\documentclass[aps,prd,twocolumn,floatfix,showpacs,superscriptaddress,showkeys,longbibliography]{revtex4-1}
\usepackage{amsmath}
\usepackage{amssymb}
\usepackage{amsfonts}
\usepackage{graphicx}
\usepackage{bbold}
\usepackage{bm}
\usepackage{bm}
\usepackage{cancel}
\usepackage{times,float}
\usepackage{graphicx}
\usepackage[usenames,dvipsnames,svgnames]{xcolor}
\usepackage{slashed}
\usepackage{hyperref}
\hypersetup{colorlinks=true, linkcolor=Blue, citecolor=PineGreen,urlcolor=Blue}
\usepackage{multirow}
\usepackage{ulem}
\usepackage{caption}
\usepackage{subcaption}

\newcommand{\be}{\begin{equation}}
\newcommand{\ee}{\end{equation}}
\newcommand{\bea}{\begin{eqnarray}}
\newcommand{\eea}{\end{eqnarray}}
\newcommand{\beas}{\begin{eqnarray*}}
\newcommand{\eeas}{\end{eqnarray*}}

\newcommand{\dpi}{(2\pi)}
\newcommand{\p}{\parallel}
\newcommand{\pp}{p_\parallel}
\newcommand{\ga}{\gamma^{\alpha}}
\newcommand{\gm}{\gamma^{\mu}}
\newcommand{\gn}{\gamma^{\nu}}

\newcommand{\ps}{\slashed{p}}
\newcommand{\rs}{\slashed{r}}

\newcommand{\ssl}{\slashed{s}}

\newcommand{\nn}{\nonumber}
\newcommand{\eB}{\left|q_f B\right|}
\newcommand{\wwp}{\omega_p}
\newcommand{\wq}{\omega_q}
\newcommand{\wk}{\omega_k}
\newcommand{\ii}{\mathrm{i}}

\newcommand{\kp}{k_\parallel}

\newcommand{\tp}{t_\parallel}

\newcommand{\tst}{\slashed{t}_\perp}

\newcommand{\rp}{r_\parallel}
\newcommand{\Sp}{s_\parallel}

\newcommand{\Qp}{Q_\parallel}

\newcommand{\lp}{\ell_\parallel}

\newcommand{\qt}{q_\perp}

\newcommand{\qp}{q_\parallel}

\newcommand{\gmp}{\gamma^\mu_\parallel}
\newcommand{\gmt}{\gamma^\mu_\perp}
\newcommand{\gnp}{\gamma^\nu_\parallel}
\newcommand{\gnt}{\gamma^\nu_\perp}
\newcommand{\gap}{\gamma^\alpha_\parallel}
\newcommand{\gat}{\gamma^\alpha_\perp}

\begin{document}

\title{Anisotropic photon emission from gluon fusion and splitting in a strong magnetic background I: The two-gluon one-photon vertex}

\author{Alejandro Ayala}
\affiliation{Instituto de Ciencias Nucleares, Universidad Nacional Aut\'onoma de M\'exico, Apartado Postal 70-543, CdMx 04510, Mexico.}
\affiliation{Centre for Theoretical and Mathematical Physics, and Department of Physics, University of Cape Town, Rondebosch 7700, South Africa.}
\affiliation{Departamento de F\'\i sica, Universidade Federal de Santa Maria, Santa Maria, RS 97105-900, Brazil.}

\author{Jorge David Casta\~no-Yepes}
\affiliation{Instituto de Física, Pontificia Universidad Católica de Chile, Vicuña Mackenna 4860, Santiago, Chile.}

\author{L. A. Hern\'andez}
\affiliation{Departamento de F\'isica, Universidad Aut\'onoma Metropolitana-Iztapalapa, Av. San Rafael Atlixco 186, C.P, CdMx 09340, Mexico.}
\affiliation{Facultad de Ciencias de la Educaci\'on, Universidad Aut\'onoma de Tlaxcala, Tlaxcala, 90000, Mexico.
}

\author{Ana Julia Mizher}
\affiliation{Instituto de F\'isica Teórica, Estadual Paulista, Rua Dr. Bento Teobaldo Ferraz, 271 - Bloco II, 01140-070 S\~ao Paulo, SP, Brazil.}
\affiliation{Centro de Ciencias Exactas, Universidad del B\'io-B\'io,
Avda. Andr\'es Bello 720, Casilla 447, 3800708, Chill\'an, Chile.}

\author{Mar\'ia Elena Tejeda-Yeomans}
\affiliation{Facultad de Ciencias - CUICBAS, Universidad de Colima, Bernal D\'iaz del Castillo No. 340, Col. Villas San Sebasti\'an, 28045 Colima, Mexico.}
\affiliation{Perimeter Institute for Theoretical Physics, 31 Caroline Street North Waterloo, Ontario N2L 2Y5, Canada.}

\author{R. Zamora}
\affiliation{Instituto de Ciencias B\'asicas, Universidad Diego Portales, Casilla 298-V, Santiago, Chile.}
\affiliation{Centro de Investigaci\'on y Desarrollo en Ciencias Aeroespaciales (CIDCA), Fuerza A\'erea de Chile, Casilla 8020744, Santiago, Chile.}
\date{\today}

\begin{abstract}

Having in mind the pre-equilibrium stage in peripheral heavy-ion collisions as a possible scenario for the production of electromagnetic radiation, we compute the two-gluon one-photon vertex in the presence of an intense magnetic field at one-loop order. The quarks in the loop are taken such that two of them occupy the lowest Landau level, with the third one occupying the first exited Landau level. When the field strength is the largest of the energy (squared) scales, the tensor basis describing this vertex corresponds to two of the three vector particles polarized in the longitudinal direction whereas the third one is polarized in the transverse direction. However, when the photon energy is of order or larger than the field strength, the explicit one-loop computation contains extra tensor structures that spoil the properties of the basis, compared to the case when the field strength is the largest of the energy scales, which signals that the calculation is incomplete. Nevertheless, by projecting the result onto the would-be basis, we show that the squared amplitude for processes involving two gluons and one photon exhibits the expected properties such as a preferred in-plane photon emission and a slightly decreasing strength for an increasing magnetic field strength. We comment on possible venues to improve the one-loop calculation that include accounting for progressive occupation of the three quarks of the lowest and first excited Landau levels such that, still working in the large field limit, a more complete description can be achieved when the photon energy increases.
\end{abstract}


\maketitle
\section{Introduction}\label{introduction}
There are several intriguing properties associated to the direct photons produced in the aftermath of relativistic heavy-ion collisions. The first is the large magnitude of their ellip\-tic flow coefficient, $v_2$, found to be similar to that of hadrons~\cite{PHENIX:2011oxq,ALICE:2018dti,PHENIX:2015igl}. Since the latter comes mainly from the late stages of the collision, when flow is already built up, it may be thought that direct photons are also preferably produced during the hadronic part of the system's evolution. However, the yields have a large thermal component that dominates over the prompt one, for low values of the transverse momentum, $p_T$. In fact, the low $p_T$ part of the spectrum is used to characterise the system's temperature which turns out to have large values that can originate only during the very early thermal history of the collision. The early emission of the bulk of the direct photons seems to be confirmed by the measured $p_T$ dependence of $v_2$ which, for large $p_T$ is consistent with zero. This can be understood when considering that photons, being a penetrating probe, can only be boosted by conditions experienced at the times when they are produced. If they in fact come from the early stages, when velocities are small, their $v_2$ for large $p_T$ should tend to zero~\cite{David:2019wpt}, as observed. Put together, these properties have been dubbed the {\it direct photon puzzle}. 

In addition, an excess of low $p_T$ photons, with respect to known sources and even to descriptions that work well for other electromagnetic probes, has been found by PHENIX~\cite{PHENIX:2014nkk}. Their analyses show that the yield of these low $p_T$ photons scales with a given power of the number of binary collisions, both in Au+Au and Cu+Cu systems~\cite{PHENIX:2018che}, which suggests that the source of these photons is similar for different colliding species and beam energies. However, it should be pointed out that a tension exists between the photon yields measured by PHENIX and STAR~\cite{STAR:2016use} and that for the latter as well as for recent ALICE measurements~\cite{ALICE:2015xmh}, the difference with state-of-the art calculations for direct photon emission~\cite{Paquet:2015lta,vanHees:2014ida,Gale:2021emg} is either not present or exists only within experimental uncertainties. Future photon measurements in a lower energy domain, such as the one to be carried out by the NICA-MPD experiment~\cite{MPD:2022qhn}, promises to provide valuable complementary information.

Attempts to find possible missing contributions for the description of the photon yield have recently payed attention to the electromagnetic radiation produced during a pre-equilibrium stage.~\cite{Monnai:2022hfs,Churchill:2020uvk}. At early times, an anisotropy in the gluon distribution, caused by a possible anisotropy in the pressure, together with a delayed equilibration of the Glasma, may lead to an anisotropic photon emission~\cite{McLerran:2014hza}. A momentum-anisotropic QGP, with a hydrodynamic evolution of the momentum distributions from the initial stage, may also contribute to photon emission and have a noticeable impact on $v_2$ for intermediate and large $p_T$~\cite{Kasmaei:2019ofu}. 

Another source of extra electromagnetic radiation, that at the same time provides a natural anisotropic emission, is the presence of a magnetic field. A magnetic field provides not only a direction that breaks rotational invariance and can be a source of $v_2$ but it also opens new channels for photon emission. For instance, in a quark-gluon plasma, photons can be emitted by magnetic field induced bremsstrahlung and pair annihilation~\cite{Zakharov:2016mmc} or by quark synchrotron radiation~\cite{Tuchin:2014pka}. The QED$\times$QCD conformal anomaly~\cite{Basar:2012bp} or fluctuations of the gluon coupled to the photon stress tensor~\cite{Basar:2014swa} can be a source of soft photons as well. Holography has also been employed to describe photon production from a strongly coupled plasma in the presence of intense magnetic fields~\cite{Avila:2022cpa,Arciniega:2013dqa,Mamo:2012kqw,Wu:2013qja}. Recall that magnetic fields of a sizeable intensity may be produced in semi-central heavy-ion collisions~\cite{Kharzeev:2007jp,Skokov:2009qp,Voronyuk:2011jd,McLerran:2013hla,Bzdak:2011yy}. Although the intensity of the field generated by spectators drops very fast, which causes an incomplete electromagnetic response in the medium formed at later times~\cite{Wang:2021oqq}, the field is found to be very intense during pre-equilibrium. Recent experimental results corroborate earlier theoretical predictions indicating a peak value $B\approx 10^{19}$ G for RHIC energies~\cite{STAR:2019wlg,Brandenburg:2021lnj}. Experimental analyses aiming to characterise the time evolution of the field are on its way. 

In a couple of recent works~\cite{Ayala:2017vex,Ayala:2019jey} we have put forward the idea that the presence of a magnetic field in the pre-equilibrium stage of the heavy-ion collision opens the gluon fusion and splitting channels for photon production. As a consequence, these magnetic field induced processes, together with the large abundance of soft pre-equilibrium gluons, contribute to enhance the photon yield and $v_2$. In these works a drastic approximation, whereby the field intensity is taken as the largest energy (squared) scale, was employed. This approximation limits the accuracy of the results to the very low $p_T$ part of the spectra. In this work we relax this approximation and do not set such restriction between the magnitude of the photon $p_T^2$ and the field intensity, although we still work in the strong field limit as compared to the square of the masses of the active quark species. Given the complexity of the calculation, here we limit ourselves to computing the one-loop two-gluon one-photon vertex that describes this process and reserve the application of its contribution to the photon yield and $v_2$ for a follow up work. It is worth mentioning that the techniques used in Refs.~\cite{Ayala:2017vex,Ayala:2019jey}, have also influenced recent treatments to obtain the corrections to the anomalous magnetic moment of the electron/muon and the corresponding ones of the anomalous magnetic moment of quarks, in the presence of a magnetic field with a strength comparable to the fermion masses~\cite{Lin:2021bqv,Xu:2020yag}.

The work is organized as follows: In Sec.~\ref{sec2} we set up the ingredients to compute the vertex function describing the on-shell coupling between two gluons and a photon in the presence of a magnetic field. In the strong field limit, we spell out the most general expression for this vertex consistent with parity, charge conjugation and gauge invariance. In Sec.\ref{sec3}, we explicitly compute the one-loop contribution to the vertex function in the presence of a magnetic field. We make use of the Landau level representation of the quark propagators and work in the strong field limit, with the quarks occupying the lowest possible Landau levels that produce a non-vanishing result. This requires that two of the quarks are in the lowest (LLL) and the other in the first excited (1LL) Landau levels. We explicitly compute the coefficients for the tensor basis that describe the matrix element and show that when the photon energy squared is allowed to be comparable to the magnetic field strength, the explicit one-loop calculation contains extra terms not present in the basis corresponding to the case where the magnetic field strength dominates the energy scales of the problem. We finally summarise and provide an outlook of the calculation to set a possible route to avoid these shortcomings in Sec.~\ref{sec4}.

\section{Two-gluon one-photon vertex in a magnetic background}\label{sec2}
The coupling between two gluons and one photon is made possible by the presence of the external magnetic field. According to Furry's theorem, the coupling vanishes in the absence of this field since  both QED and QCD are charge conjugation conserving theories. The breaking of Lorentz invariance, also induced by the magnetic field, produces that space-time is separated into parallel and perpendicular pieces, with respect the magnetic field. For definiteness, let us consider a constant in time and spatially uniform magnetic field of strength $B$ pointing along the $\hat{z}$ direction. The separation of space-time is implemented by introducing the tensor metric components such that~\cite{PhysRevD.101.036016}
\begin{subequations}
\bea
g^{\mu\nu} = g_{\parallel}^{\mu\nu} + g_{\perp}^{\mu\nu},
\eea
\bea
g_{\parallel}^{\mu\nu} &=& \text{diag}(1,0,0,-1),\nonumber\\
g_{\perp}^{\mu\nu} &=& \text{diag}(0,-1,-1,0),
\eea
\end{subequations}
which implies that for any four-vector $p^{\mu}$, we can write
\begin{subequations}
\begin{eqnarray}
p_\parallel^\mu&=&(p_0,0,0,p_3),\nn\\
p_\perp^\mu&=&(0,p_1,p_2,0)
\end{eqnarray}
and
\begin{eqnarray}
p^2 = p_{\parallel}^2 - p_{\perp}^2,
\end{eqnarray}
\end{subequations}
where $p_\parallel^2\equiv p_0^2-p_3^2$ and  $p_{\perp}^2=p_1^2+p_2^2$. Therefore, the most general tensor structure for a third rank tensor, such as the two-gluon one-photon vertex, $\Gamma^{\mu\nu\alpha}_{ab}$, where $\mu,\nu,\alpha$ and $a,b$ are Lorentz and color indices, respectively, involves the metric tensors in the parallel and perpendicular directions
\bea
g_{\parallel}^{\mu\nu},~g_{\perp}^{\mu\nu}
\eea
and the momentum components of the gluons and the photon, also in the parallel and perpendicular directions, namely,
\bea
p^\alpha_\parallel,~p^\alpha_\perp,~k^\alpha_\parallel,~k^\alpha_\perp,~q^\alpha_\parallel,~q^\alpha_\perp,
\eea
where $p$, $k$ and $q$ are the four-momenta of the gluons and of the photon, respectively.

The kind of magnetic field hereby considered cannot transfer energy-momentum to the gluons and the photon. Thus, when also neglecting possibly medium induced modification on their dispersion properties, energy-momentum conservation imposes that for on-shell propagation all the four-momenta are parallel~\cite{Adler:1970gg}
\bea
q^\mu=p^\mu+k^\mu.
\eea
Choosing $q^\mu$ as the reference four-momentum, we have
\begin{subequations}
\bea
p^\mu=\left(\frac{\omega_p}{\omega_q}\right)q^\mu,
\eea
\bea
k^\mu=\left(\frac{\omega_k}{\omega_q}\right)q^\mu,
\eea
\label{colinealmomenta}
\end{subequations}
where $\omega_p,~\omega_k,~\omega_q$ are the energies of the gluons and photon, respectively. The on-shell restriction implies a reduction of the tensor structures involved in the vertex construction. Then, it is enough to consider tensors obtained from the combination of
\bea
g_{\parallel}^{\mu\nu},~g_{\perp}^{\mu\nu},~q^\alpha_\parallel,~q^\alpha_\perp.
\eea

Following the findings of Refs.~\cite{Ayala:2017vex,Ayala:2019jey}, in the approximation where the magnetic field is the largest of the kinematical energy (squared) variables, the tensor structure can be expressed as 
\bea
\Gamma^{\mu\nu\alpha}_{ab}&=&\delta_{ab}\Gamma^{\mu\nu\alpha},\nn\\
\Gamma^{\mu\nu\alpha}&\equiv&\Gamma_1(\omega_q,\omega_k,q^2)\frac{\epsilon_{ij} q_\perp^i g_\perp^{j\mu}}{\sqrt{q_\perp^2}}\left(g_\parallel^{\nu\alpha}-\frac{q_\parallel^\nu q_\parallel^\alpha}{q_\parallel^2}\right)\nn\\
&+&\Gamma_2(\omega_q,\omega_k,q^2)\frac{\epsilon_{ij} q_\perp^i g_\perp^{j\nu}}{\sqrt{q_\perp^2}}\left(g_\parallel^{\mu\alpha}-\frac{q_\parallel^\mu q_\parallel^\alpha}{q_\parallel^2}\right)\nn\\
&+&\Gamma_3(\omega_q,\omega_k,q^2)\frac{\epsilon_{ij} q_\perp^i g_\perp^{j\alpha}}{\sqrt{q_\perp^2}}\left(g_\parallel^{\mu\nu}-\frac{q_\parallel^\mu q_\parallel^\nu}{q_\parallel^2}\right)\nonumber\\
&\equiv&\sum_{n=1}^3 \Gamma_n(\omega_q,\omega_k,q^2)\Gamma^{\mu\nu\alpha}_n,
\label{vertex}
\eea
where $\Gamma_n(\omega_q,\omega_k,q^2)$, $n=1,2,3$, are scalar coefficients and $\epsilon_{ij}$ is the Levy-Civita symbol in the transverse components: $\epsilon_{12}=-\epsilon_{21}=1$. Since $q^2=0$, we can use either $q_\parallel^2$ or $q_\perp^2$ to describe the functional dependence of the coefficients $\Gamma_n$ on the photon's momentum. Notice that
\bea
v^\beta_\perp\equiv\epsilon_{ij} \frac{q_\perp^i g_\perp^{j\beta}}{\sqrt{q_\perp^2}}
\label{vperp}
\eea
corresponds to the polarization vector for transverse (with respect to the magnetic field) modes and also that in the strong field limit, two of the three particles, either one of the gluons and the photon or the two gluons, propagate in the parallel polarization mode, characterised by a vector $v^\sigma_\parallel$ that satisfies~\cite{PhysRevD.97.014023}
\bea
v^\sigma_\parallel v^\rho_\parallel\equiv \Pi^{\sigma\rho}_\parallel=g_\parallel^{\sigma\rho}-\frac{q_\parallel^\sigma q_\parallel^\rho}{q_\parallel^2}.
\label{pipar0}
\eea
The tensor basis of Eq.~(\ref{vertex}) is orthonormal, namely,
\bea
   \Gamma^{\sigma\rho\beta}_n\Gamma_{\sigma\rho\beta m}=\delta_{nm}.
\label{ortogonalidad}
\eea
This basis is also complete, when the magnetic field can be taken as the largest possible energy (squared). Relaxing this condition introduces extra terms in the basis whose importance increases as the ratio $\omega_q^2/|eB|$ increases.

\begin{figure}[b]
    \centering
    \includegraphics[scale=0.35]{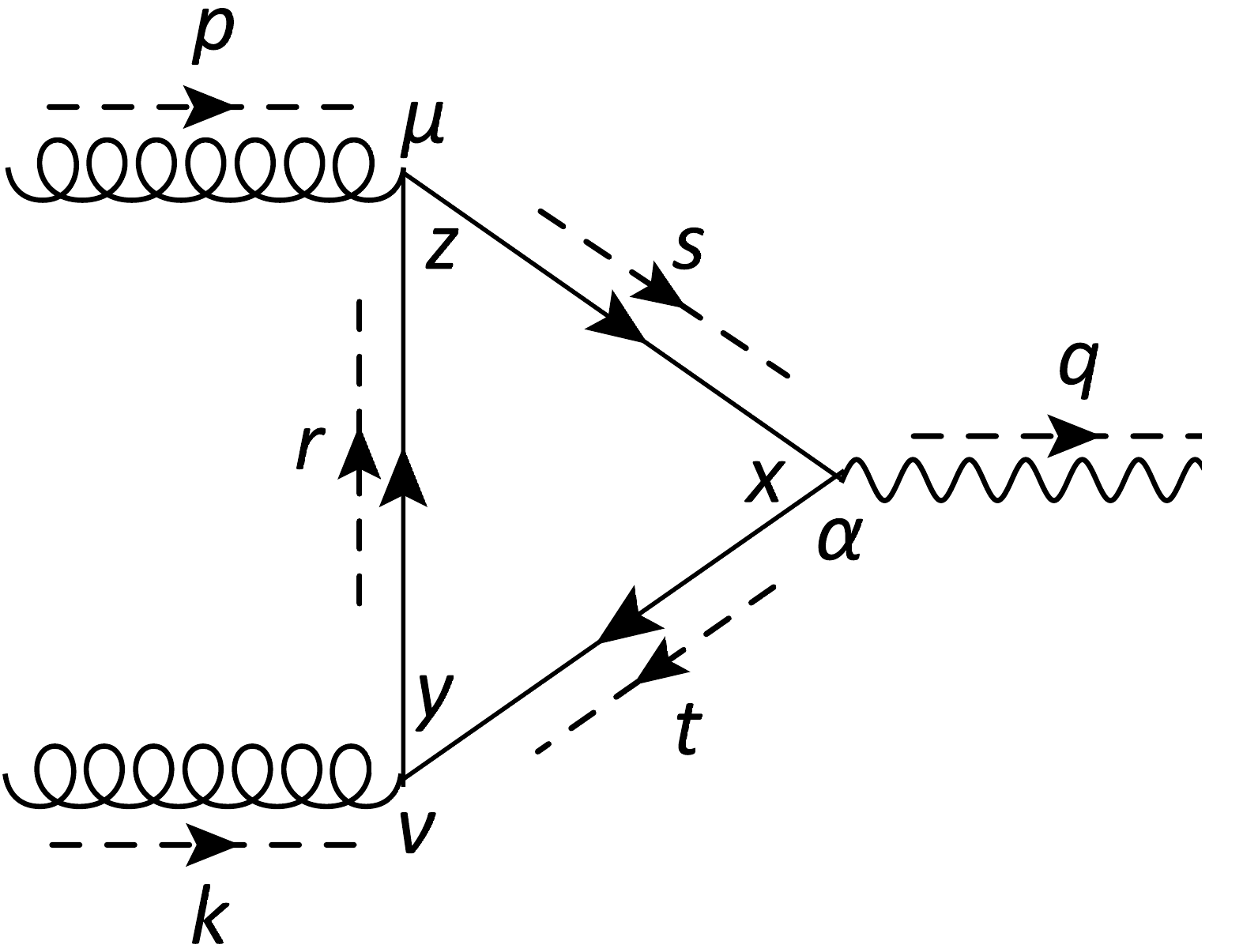}
    \caption{Feynman diagram describing the coupling between two gluons and a photon at the lowest order in $\alpha_s$ and $\alpha_\text{em}$ in the presence of a magnetic field.}
    \label{fig:Fusion}
\end{figure}
In the strong field approximation, the physical picture that emerges is as follows: It is well known that a strong magnetic field forces two of the vector particles to occupy parallel polarization states~\cite{PhysRevD.101.036016,PhysRevD.83.111501,Ayala:2018ina}. When the vertex involves a third vector particle, invariance under charge conjugation and conservation of angular momentum require that its polarization state is transverse. At the lowest perturbative order, this can be understood recalling that, when polarized in the same direction, the addition of the three spin 1/2 quarks in the loop gives rise to a half-integer spin state that cannot describe the spin state of a combination of three vector particles. Therefore, one of the quarks that make up the loop needs to be placed not in the LLL but instead in the 1LL. This in turn induces the emergence of a transverse mode to be occupied by one of the vector particles. Similar selection rules, albeit in the weak field limit, are discussed in Ref.~\cite{Adler:1970gg}.
\begin{figure}[b]
    \centering
    \includegraphics[scale=0.25]{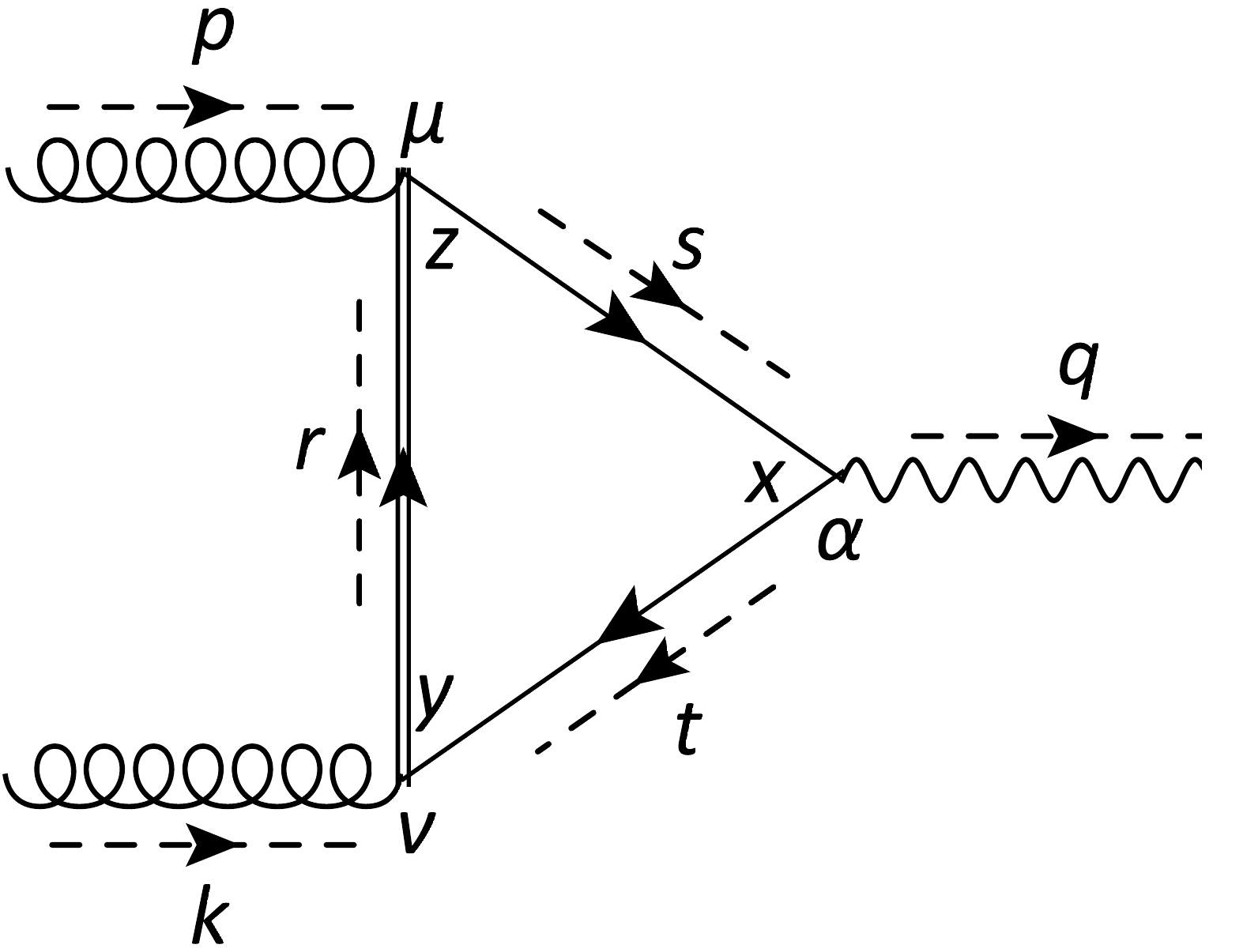}\includegraphics[scale=0.25]{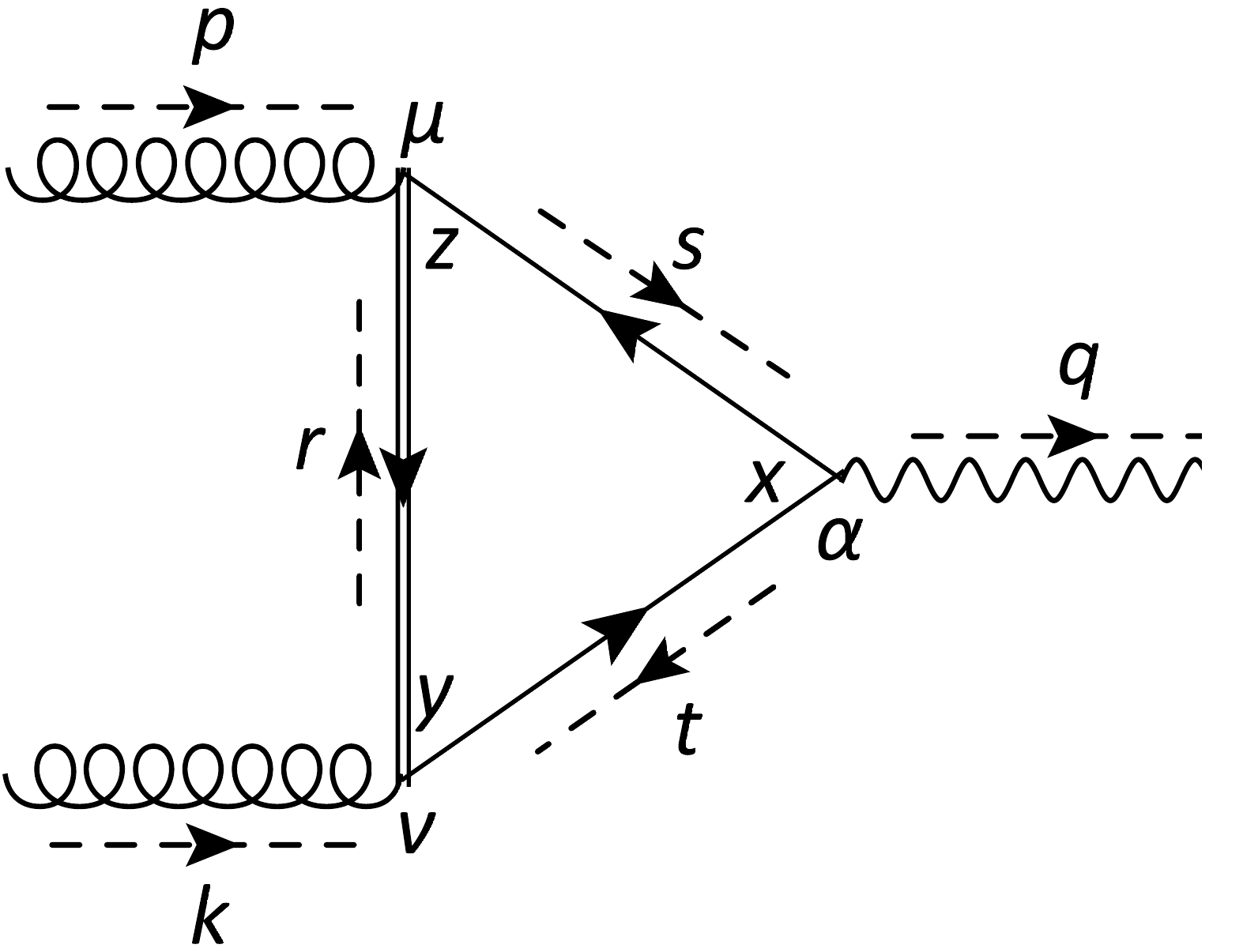}\\
    \vspace{0.3cm}
    \includegraphics[scale=0.25]{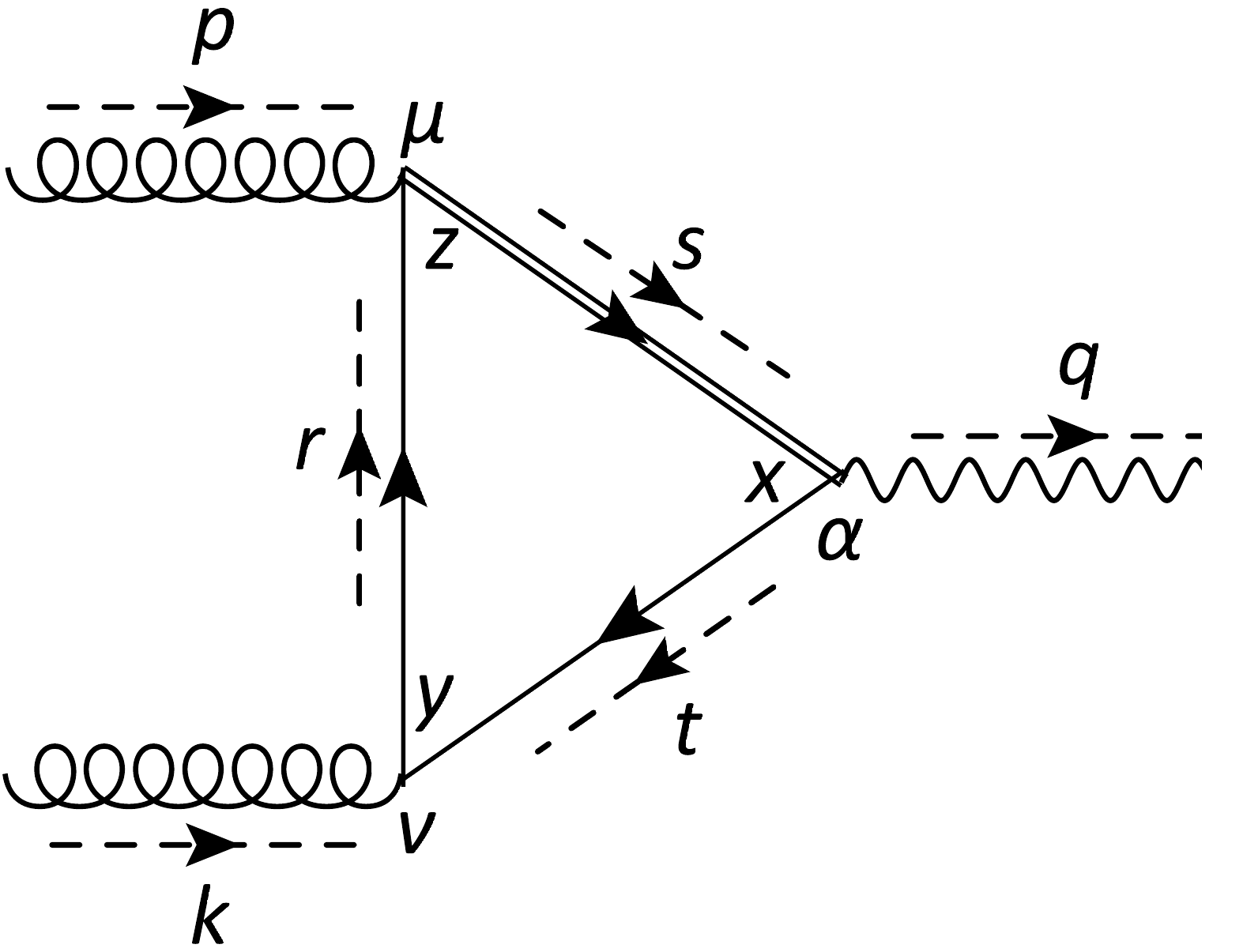}\includegraphics[scale=0.25]{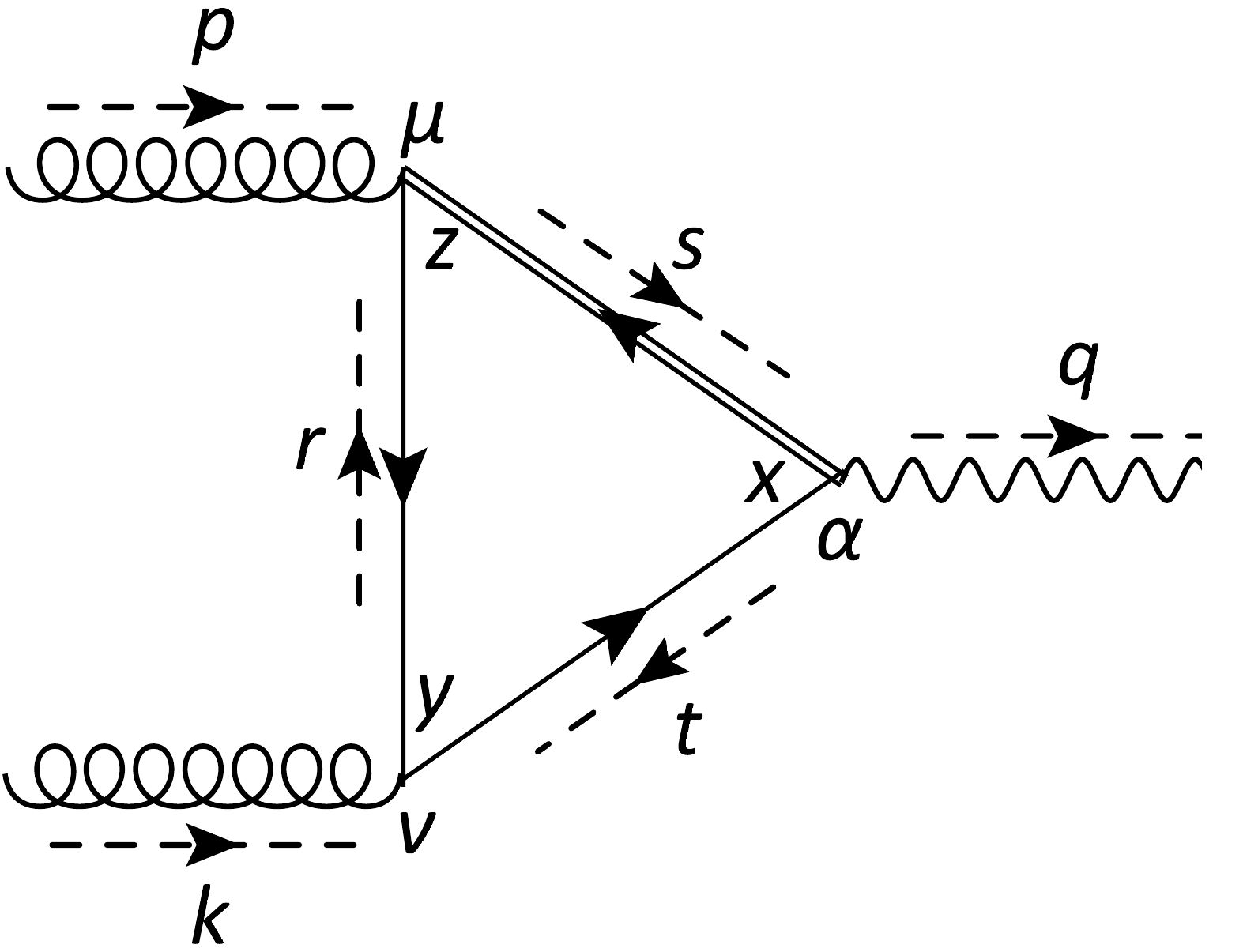}\\
    \vspace{0.3cm}
    \includegraphics[scale=0.25]{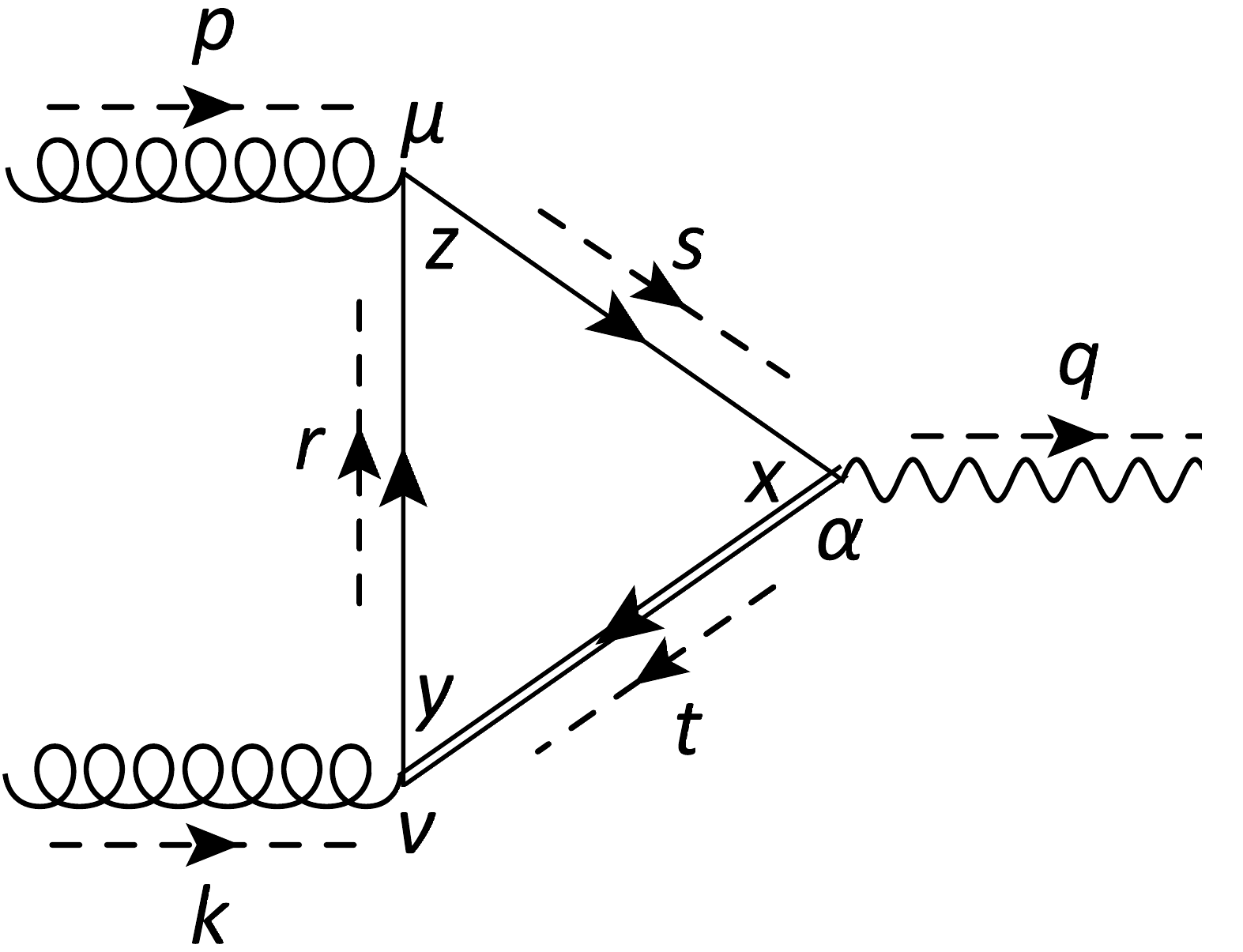}\includegraphics[scale=0.25]{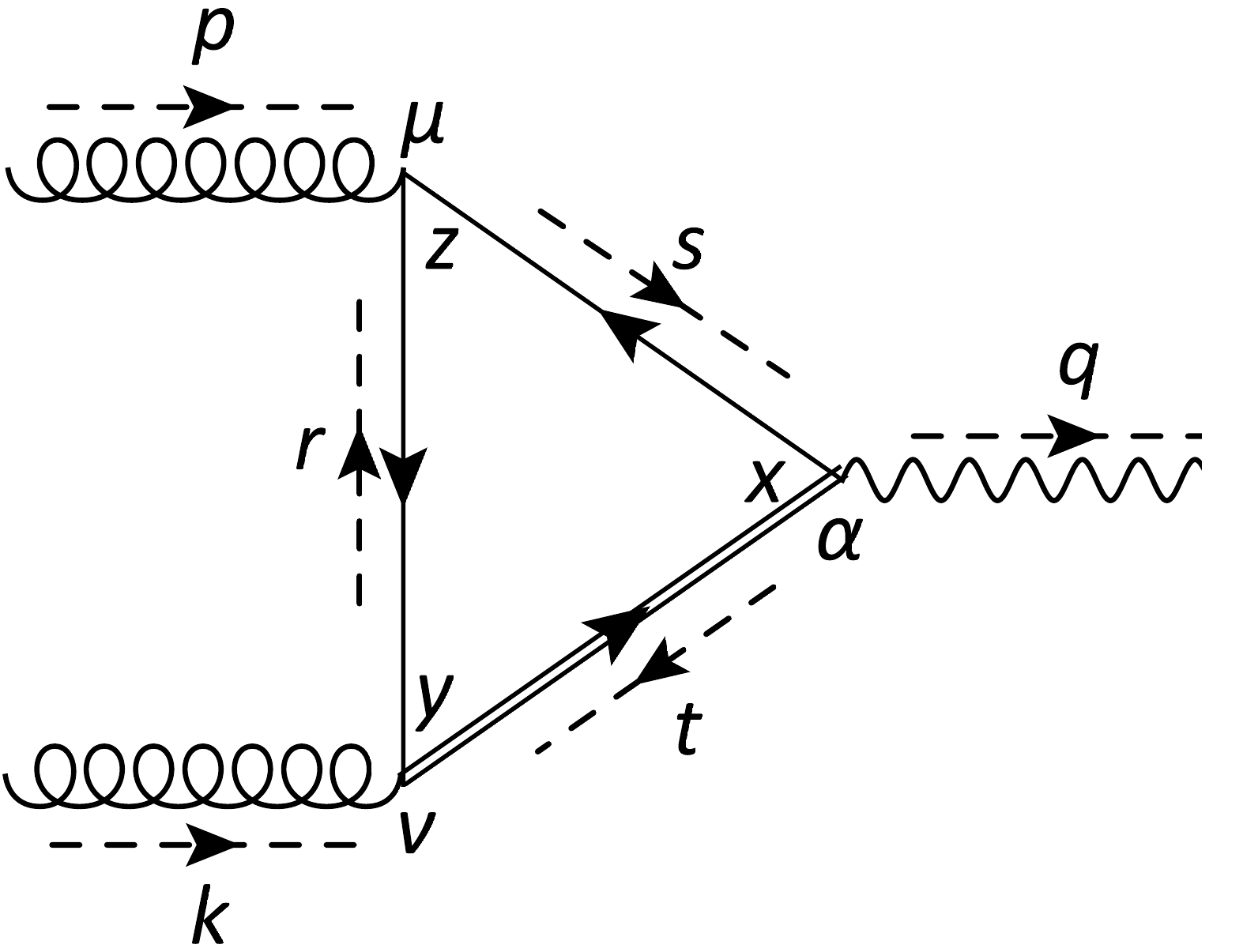}
    \caption{Leading order Feynman diagrams describing the vertex coupling two gluons and one photon in the presence of a magnetic field. The single lines represent fermion propagators in the lowest Landau level, and the double lines are propagators in the first excited Landau level. The continuous and dashed arrows represent the flow of charge and momentum, respectively.}
    \label{fig:TrianglesFusion}
\end{figure}

Parity conservation requires that the vertex is symmetric under the exchange of the gluon Lorentz indices $\mu\leftrightarrow\nu$ which in turn requires that when $\omega_p\leftrightarrow\omega_k$,  $\Gamma_1\leftrightarrow\Gamma_2$, while $\Gamma_3$ remains invariant. The vertex also satisfies the transversality condition
\bea
q_\mu\Gamma^{\mu\nu\alpha} = q_\nu \Gamma^{\mu\nu\alpha}=q_\alpha\Gamma^{\mu\nu\alpha}=0,
\label{transversality}
\eea
imposed by gauge invariance. Relaxing the strong field approximation to partially account for contributions to allow that the photon energy (squared) is not small compared to the magnetic field intensity, spoils the symmetry properties. We explicitly show this in the following section, where we compute the coefficients $\Gamma_n$ at leading order in the strong magnetic field at one-loop level. This calculation provides the key features of photon production in the kinematical regions of interest in the context of the \textit{direct photon puzzle} and sheds light on the road to improve the approximation.

\section{One-loop vertex in the strong field limit}\label{sec3}
The scattering process involving two gluons and a photon, either gluon fusion or splitting, is described at leading order in the strong, $\alpha_s=g^2/4\pi$, and electromagnetic, $\alpha_\text{em}=e^2/4\pi$, couplings by a Feynman diagram that is represented as a fermion triangle with two gluons and one photon attached to the vertices. Given that the magnetic field breaks Lorentz symmetry, the vertex describing these processes needs to be computed starting from configuration space. Figure~\ref{fig:Fusion} shows the generic diagram where the internal lines correspond to fermion propagators in the presence of a magnetic field, which can be written as
\bea
S(x,x')=\Phi(x,x')\int\frac{d^4p}{(2\pi)^4}e^{-\ii p\cdot(x-x')}S(p),
\eea
where $\Phi(x,x')$ is Schwinger's phase factor that, for a fermion with charge $q_f$, is given by
\bea
\!\!\!\!\!\!\!\!\Phi(x,x')=\exp\left\{\ii q_f\int_{x'}^x\! d\xi^\mu\left[A^\mu+\frac{1}{2}F_{\mu\nu}(\xi-x')^\nu\right]\right\}\!.
\label{schwingerfactor}
\eea

The Fourier transform of the translationally invariant part of the propagator can be expressed as a sum over Landau levels, such that
\bea
\!\!\!\!\!\!\!\ii S(p)=\ii e^{-p_\perp^2/\eB}\sum_{n=0}^{+\infty}\frac{(-1)^nD_n(|q_fB|,p)}{p_\parallel^2-m_f^2-2n\left |  q_fB\right |+\ii\epsilon},
\label{SLandauLevels}
\eea
with $m_f$ the fermion mass and
\bea
D_n(|q_fB|,p)&=&2(\slashed{p}_\parallel+m_f)\mathcal{O}^{-}L_n^0\left(\frac{2p_\perp^2}{\left | q_fB \right|}\right)\nn\\
&-&2(\slashed{p}_\parallel+m_f)\mathcal{O}^{+}L_{n-1}^0\left(\frac{2p_\perp^2}{\left | q_fB \right |}\right)\nn\\
&+&4\slashed{p}_\perp L_{n-1}^1\left(\frac{2p_\perp^2}{\left | q_fB \right |}\right),
\label{Dn}
\eea
where $L_n^m(x)$ are the generalized Laguerre polynomials, and the operators $\mathcal{O}^{\pm}$ are given by
\bea
\mathcal{O}^{\pm}=\frac{1}{2}\left[1\pm \ii\gamma^1\gamma^2\text{sign}(q_f B)\right].
\label{Op}
\eea

From the Feynman diagram of Fig.~\ref{fig:Fusion}, we write the expression for the vertex as
\bea
   \Gamma_{\mu\nu\alpha}^{ab}&=&-\int \!d^4xd^4yd^4z\int\!\frac{d^4r}{(2\pi)^4}
   \frac{d^4s}{(2\pi)^4}\frac{d^4t}{(2\pi)^4}\nn\\
   &\times&e^{-\ii t\cdot (y-x)}e^{-\ii s\cdot (x-z)}e^{-\ii r\cdot (z-y)}e^{-\ii p\cdot z}e^{-\ii k\cdot y}e^{\ii q\cdot x}\nonumber\\
   &\times&
   \Big\{
   {\mbox{Tr}}\left[ \ii q_f\gamma_\alpha \ii S(s) \ii g\gamma_\mu t^a \ii S(r) ig\gamma_\nu t^b \ii S(t) \right]\nn\\
   &+&
   {\mbox{Tr}}\left[ \ii q_f\gamma_\alpha \ii S(t) \ii g\gamma_\nu t^b \ii S(r) ig\gamma_\mu t^a \ii S(s) \right]
   \Big\}
   \nonumber\\
   &\times&\Phi(x,y)\Phi(y,z)\Phi(z,x),
   \label{amplitude}
\eea
where $p$ and $k$ are the gluon and $q$ the photon four-mo\-men\-ta, $t^a=\lambda^a/2,\ t^b=\lambda^b/2$ with $\lambda^a$ and $\lambda^b$ being Gell-Mann matrices. Notice that in the last equation the contribution from the charge-conjugate diagram is also considered. 

To describe a constant magnetic field that points in the $z$-direction, the vector potential $A^\mu$ can be chosen in the symmetric gauge,
\bea
A^\mu=\frac{B}{2}(0,-y,x,0),
\eea
so that, from Eq.~(\ref{schwingerfactor}), the product of Schwinger phases is given by
\bea
\!\!\!\Phi(x,y)\Phi(y,z)\Phi(z,x)=e^{\textcolor{red}{-}\ii\frac{\eB}{2}\epsilon_{mj}(z-x)_m(x-y)_j},
\eea
where the indices $m,j=1,2$. It has been shown that when considering localised interactions, such as in the present calculation, the product of Schwinger phases is gauge invariant~\cite{Ayala:2020muk}.  

Given that at the early stages of a heavy-ion collision the magnetic field is at its peak intensity, we can work in the strong field approximation, whereby fermions in the loop occupy the lowest possible Landau levels, and one takes $|q_fB|\gg m_f^2$. In Refs.~\cite{Ayala:2017vex,Ayala:2019jey}, it has been shown that the first order non-vanishing contribution in the large field limit requires two of the fermion propagators to occupy the LLL whereas the third one occupies the 1LL, which correspond to $n=0$ and $n=1$ in Eq.~(\ref{SLandauLevels}), respectively
\bea
\ii S_\text{LLL}(p)&=&\ii\frac{e^{-p_\perp^2/q_fB}}{p_\p^2-m_f^2+\ii\epsilon}(\slashed{p}_\parallel+m_f)\mathcal{O}^{-},
\label{LLL}
\eea
\begin{widetext}
\bea
\ii S_\text{1LL}(p)=-2\ii\frac{e^{-p_\perp^2/q_fB}}{p_\p^2-m_f^2-2\left |q_fB\right |+\ii\epsilon}\left[(\slashed{p}_\parallel+m_f)\left(1-\frac{2p_\perp^2}{\left | q_fB \right |}\right)\mathcal{O}^{-}-(\slashed{p}_\parallel+m_f)\mathcal{O}^{+}+2\ps_\perp\right]\!.\nn\\
\label{propagadores2}
\eea

The contributing Feynman diagrams, obtained when placing two fermion propagators in the LLL and the other in the 1LL, are depicted in Fig.~\ref{fig:TrianglesFusion}. After integration of the configuration space variables, the vertex in Eq.~(\ref{amplitude}) is given by

\bea
\Gamma^{\mu\nu\alpha}_{ab}=-\delta^{(4)}(q-k-p)\text{Tr}\left[t_at_b\right]\frac{8\pi^4 q_f g^2}{\eB}\qp^2e^{f\left(p_{\perp}, k_{\perp}\right)}\sum_{i=1}^3D_{i}^{\mu\nu\alpha},
\label{matrixelem}
\eea
where

\bea
f\left(p_\perp,k_\perp\right)&=&\frac{1}{8\eB}\left(p_m-k_m+i\epsilon_{mj}(p_j+k_j)\right)^2\-\frac{1}{2\eB}\left(p_m^2+k_m^2+2i\epsilon_{jm}p_mk_j\right),
\eea
and

\bea
D_{1}^{\mu\nu\alpha}&=&\bigg\{2\left|\qt\right|\,\mathcal{I}_1\left[\left(\widetilde{C}v^\nu_\perp+\ii \qt^\nu\right) \left(\Pi_\parallel^{\mu\alpha}-\frac{1}{2}g_\parallel^{\mu\alpha}\right)-\left(\widetilde{C}v^\mu_\perp+\ii\qt^\mu\right)\left(\Pi_\parallel^{\nu\alpha}-\frac{1}{2}g_\parallel^{\nu\alpha}\right)\right]+2\mathcal{J}_1\epsilon_{ij}g_\perp^{i\mu}g_\perp^{j\nu}q_\parallel^\alpha\bigg\},
\label{TensorD1}
\eea

\bea
D_{2}^{\mu\nu\alpha}&=&\bigg\{2\left|\qt\right|\,\mathcal{I}_2\left[\left(\widetilde{B}v^\mu_\perp+\frac{3\ii\wk}{4\wq} \qt^\mu\right) \left(\Pi_\parallel^{\nu\alpha}-\frac{1}{2}g_\parallel^{\nu\alpha}\right)-\left(\widetilde{B}v^\alpha_\perp+\frac{3\ii\wk}{4\wq}\qt^\alpha\right)\left(\Pi_\parallel^{\mu\nu}-\frac{1}{2}g_\parallel^{\mu\nu}\right)\right]+2\mathcal{J}_2\epsilon_{ij}g_\perp^{i\alpha}g_\perp^{j\mu}q_\parallel^\nu\bigg\},\nn\\
\label{TensorD2}
\eea
\bea
D_{3}^{\mu\nu\alpha}&=&\bigg\{2\,\mathcal{I}_3\bigg[\left(-\ii\tilde{A}_1q_\perp^\nu -\tilde{A}_2\left|\qt\right|v_\perp^\nu \right)\left(\Pi_\parallel^{\mu\alpha}-\frac{1}{2}g^{\mu\alpha}_\parallel\right)+\left(\ii\tilde{A}_1q_\perp^\alpha +\tilde{A}_2\left|\qt\right|v_\perp^\alpha \right)\left(\Pi_\parallel^{\mu\nu}-\frac{1}{2}g^{\mu\nu}_\parallel\right)\Bigg]+2\mathcal{J}_3\epsilon_{ij}g_\perp^{i\nu}g_\perp^{j\alpha}q_\parallel^\mu\Bigg\},
\nonumber\\
\label{TensorD3}
\eea
\end{widetext}
with the coefficients $\widetilde{A}$, $\widetilde{B}$ and $\widetilde{C}$ are given by
\begin{subequations}
\bea
\widetilde{A}_1(\omega_k,\omega_q)\equiv-\frac{\tilde{C}}{8}-\frac{\omega_p}{\omega_q}-\frac{1}{8},
\eea

\bea
 \widetilde{A}_2(\omega_k,\omega_q)\equiv-\frac{\tilde{C}}{8}+\frac{\omega_k}{2\omega_q}+\frac{1}{4},
\eea

\bea
\widetilde{B}\equiv\frac{2\wk-3\wq}{\wq}+\frac{\ii}{2},
\eea

\bea
\widetilde{C}\equiv\frac{\wwp-\wk}{\wq},
\eea
\end{subequations}
and where we have defined
\begin{widetext}
\begin{subequations}
\bea
\mathcal{I}_1
\equiv 
\int_0^1dx\int_0^{1-x}dy\frac{x+\frac{\omega_k}{\omega_q}(1-x-y)-\left(x+\frac{\omega_k}{\omega_q}(1-x-y)\right)^2}{\Delta_1^2(x,y)},
\eea
\bea
\mathcal{J}_1&\equiv&\int_0^1dx\int_0^{1-x}dy\left[\frac{2x+2(1-x-y)\frac{\wk}{\wq}-1}{\Delta_1(x,y)\qp^2}+\frac{1}{\Delta_1^2(x,y)}\left\{\left[\left(x+\frac{\wk}{\wq}(1-x-y)\right)^2\right.\right.\right.\nn\\
&-&\left.\left.\left.\left(x+(1-x-y)\frac{\wk}{\wq}\right)\frac{\wk}{\wq}\right]+\left(x+(1-x-y)\frac{\wk}{\wq}\right)\frac{\wk(\wq-\wk)}{\wq^2}-\left[\left(x+(1-x-y)\frac{\wk}{\wq}\right)^2\right.\right.\right.\nn\\
&+&\left.\left.\left.\left(1-2x-2(1-x-y)\frac{\wk}{\wq}\right)\frac{\wk}{\wq}\right.\Bigg]\left(x+(1-x-y)\frac{\wk}{\wq}\right)\right.\Bigg\}\right.\Bigg],
\eea
\bea
\Delta_1(x,y)&=&\left[x+(1-x-y)(\wk/\wq)\right]^2\qp^2-x\qp^2-(1-x-y)\left[(\wk/\wq)^2\qp^2-2\eB\right]+m_f^2.
\label{Delta1}
\eea
\end{subequations}
\begin{subequations}
\bea
\mathcal{I}_2
\equiv
\int_0^1dx\int_0^{1-x}dy\frac{\left[\left(\frac{\wk}{\wq}x+(1-x-y)\right)^2-\frac{\wk}{\wq}\left(\frac{\wk}{\wq}x+(1-x-y)\right)\right]}{\Delta_2^2(x,y)},
\eea
\bea
\mathcal{J}_2&\equiv&\int_0^1dx\int_0^{1-x}dy\left[\frac{(1-2x)\frac{\wk}{\wq}-2(1-x-y)}{\Delta_2(x,y)\qp^2}+\frac{1}{\Delta_2^2(x,y)}\left\{\left[\left((1-x-y)-\frac{\wk}{\wq}x\right)\frac{\wk}{\wq}\right.\right.\right.\nn\\
&-&\left.\left.\left.\left((1-x-y)+\frac{\wk}{\wq}x\right)^2\right]+\left((1-x-y)+\frac{\wk}{\wq}x\right)\frac{\wk(\wq-\wk)}{\wq^2}+\left[\left((1-x-y)+\frac{\wk}{\wq}x\right)^2\right.\right.\right.\nn\\
&-&\left.\left.\left.\left(2x-1+2(1-x-y)\frac{\wk}{\wq}\right)\frac{\wk}{\wq}\right.\Bigg]\left((1-x-y)+\frac{\wk}{\wq}x\right)\right.\Bigg\}\right.\Bigg],
\eea
\bea
\Delta_2(x,y)&=&\left[(1-x-y)+(\wk/\wq)x\right]^2\qp^2-x(\wk/\wq)^2\qp^2-(1-x-y)\left(\qp^2-2\eB\right)+m_f^2.
\label{Delta2}
\eea
\end{subequations}

\bea
\mathcal{I}_3
\equiv 
\int_0^1dx\int_0^{1-x}dy \frac{-x+\frac{\omega_k}{\omega_q}\left(y-\frac{\omega_k}{\omega_q}(1-x-y)\right)+\left(x+\frac{\omega_k}{\omega_q}(1-x-y)\right)^2}{\Delta_3^2(x,y)}
\eea

\bea 
\mathcal{J}_3\equiv
\int_0^1dx\int_0^{1-x}dy\ \frac{-i}{8\pi}\frac{1}{\Delta_3^2} \left[\left(\frac{\omega_p+2\omega_k}{\omega_q}\right)\Delta_3 + \left[\tilde{f}^3-\tilde{f}^2\left(1+\frac{\omega_k}{\omega_q}\right)+\tilde{f}\frac{\omega_k}{\omega_q}\right]q_\parallel^2\right]
\eea

\begin{equation}
    \Delta_3(x,y)=\left[x+(1-x-y)\frac{\omega_k}{\omega_q}\right]^2\qp^2-x\qp^2-(1-x-y)\kp^2+2\eB y+m_f^2,
\end{equation}

\end{widetext}
with
\be
\tilde{f}\equiv x+(1-x-y)\frac{\omega_k}{\omega_q}.
\ee 

We emphasize that Eqs.~(\ref{TensorD1}) -- (\ref{TensorD3}) are obtained under the approximation whereby $\eB\gg m_f^2$ in the numerator of Eqs.~(\ref{LLL}) and~(\ref{propagadores2}). This is a reasonable approximation when accounting only for the active quark species $u$, $d$, $s$ and during the pre-equilibrium and/or very early stages of the collision. On the other hand, the masses in the denominator need to be kept finite so that the integrand is infrared safe. This is explicitly shown in Appendix~\ref{App1}.

\begin{figure*}
     \centering
     \begin{subfigure}[b]{0.32\textwidth}
         \centering
         \includegraphics[width=\textwidth]{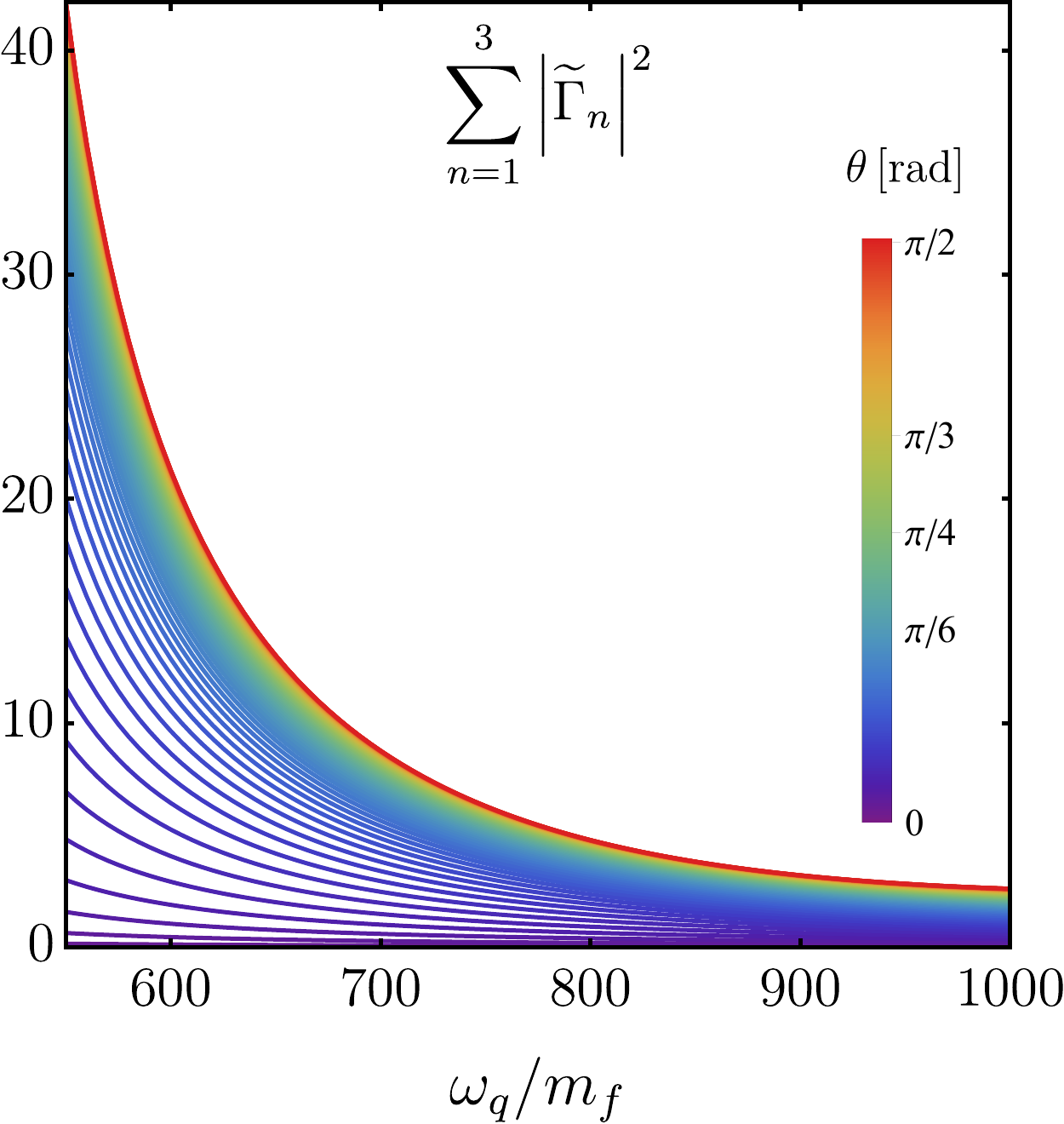}
         \caption{}
         \label{fig:GammaSquared_vs_wq_theta}
     \end{subfigure}
     \hfill
     \begin{subfigure}[b]{0.32\textwidth}
         \centering
         \includegraphics[width=\textwidth]{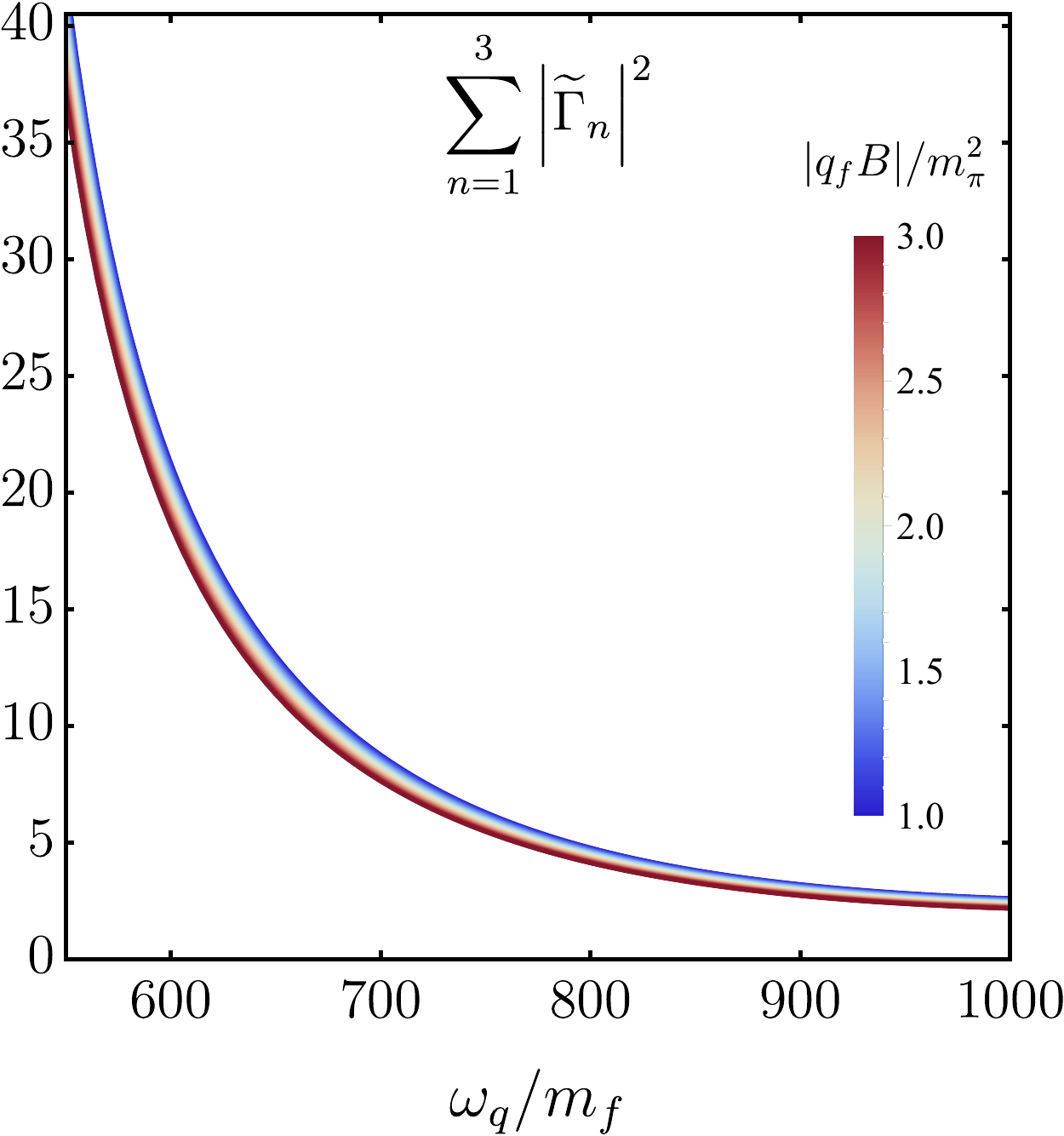}
         \caption{}
         \label{fig:GammaSquared_vs_wq_eB}
     \end{subfigure}
     \hfill
     \begin{subfigure}[b]{0.31\textwidth}
         \centering
         \includegraphics[width=\textwidth]{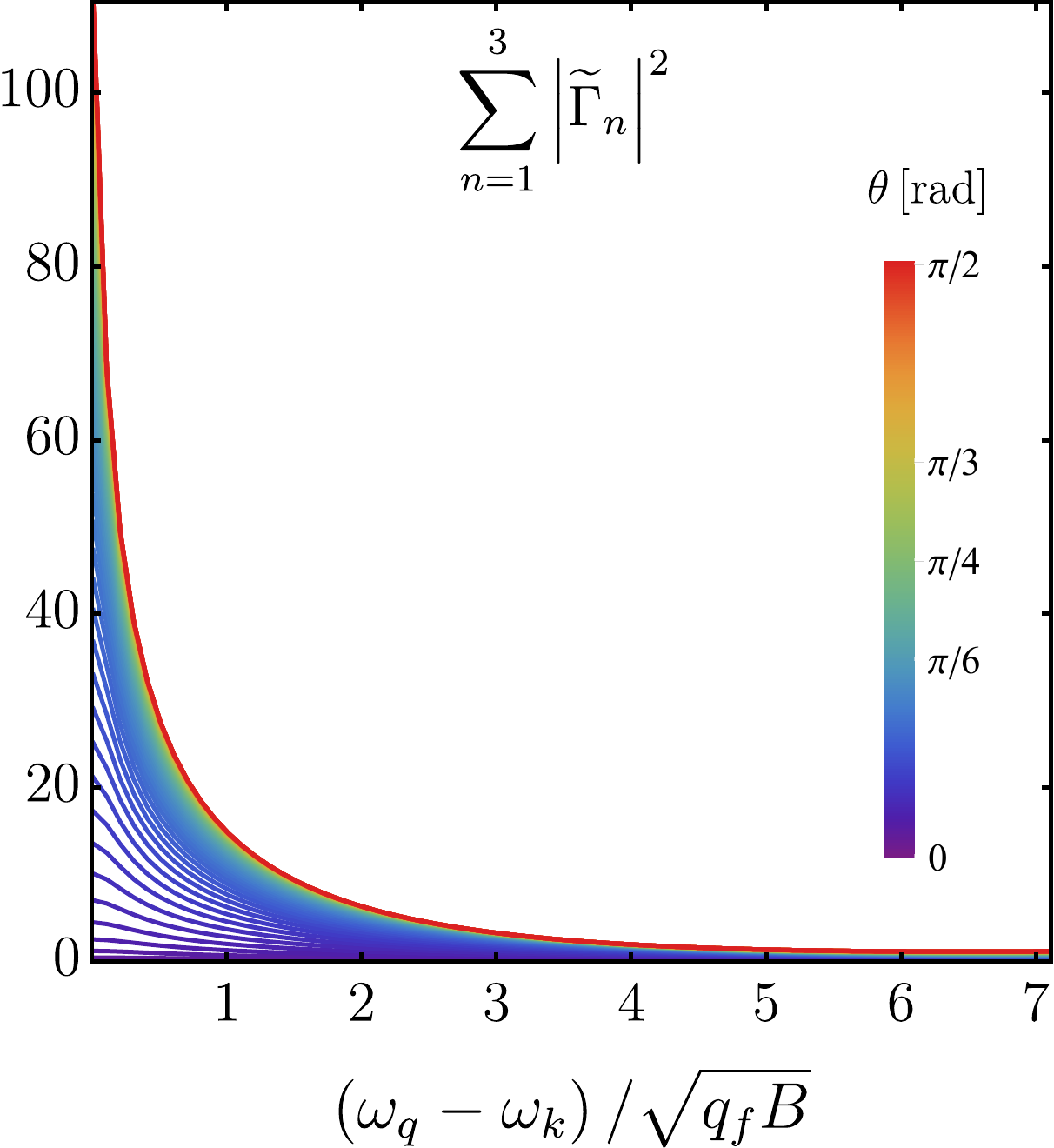}
         \caption{}
         \label{fig:GammaSquared_vs_wq_z}
     \end{subfigure}
        \caption{Sum of squared amplitudes of $\widetilde{\Gamma}_n$, $n=1,2,3$ from Eqs.~(\ref{eq:Gamma1}),~(\ref{eq:Gamma2}) and~(\ref{eq:Gamma3}), as a function of: (a) the photon energy $\omega_q$ and the angle with respect to the magnetic field $\theta$ with $\eB=m_\pi^2$, (b) the photon energy $\omega_q$ and the magnetic field strength $\eB$ with $\theta=\pi/2$, and (c) the ratio $(\wq-\wk)/\sqrt{\eB}$ and the angle with respect to the magnetic field $\theta$ with $\eB=m_\pi^2$. All the physical parameters are normalized with the quarks mass $m_f=2\times10^{-3}$ GeV, having fixed $\omega_k=1$ GeV.}
        \label{fig:three graphs}
\end{figure*}

\begin{figure*}
     \centering
     \begin{subfigure}[b]{0.32\textwidth}
         \centering
         \includegraphics[width=\textwidth]{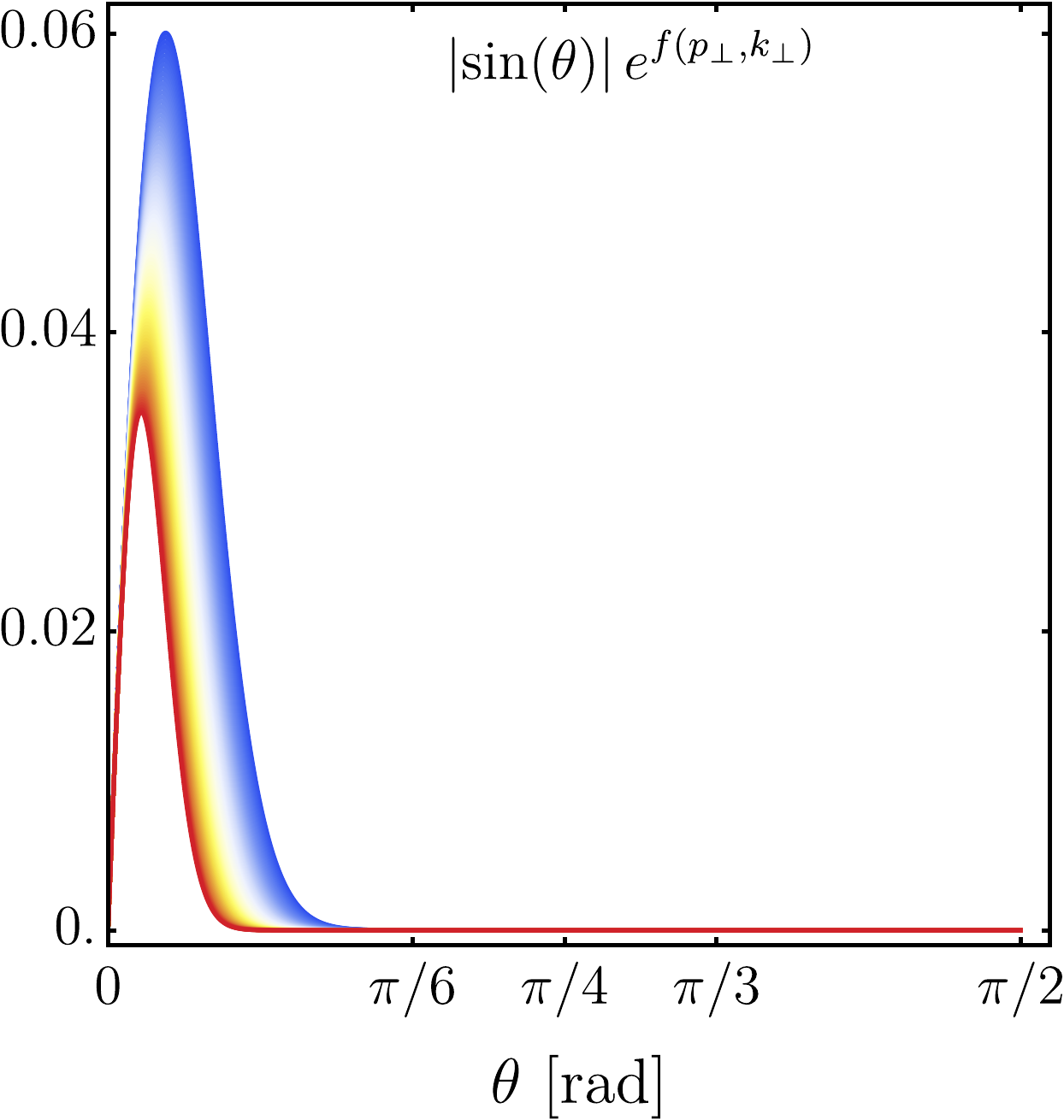}
         \caption{$\wk=1$ GeV, $\eB=m_\pi^2$}
         \label{fig:expf_a}
     \end{subfigure}
     \hfill
     \begin{subfigure}[b]{0.32\textwidth}
         \centering
         \includegraphics[width=\textwidth]{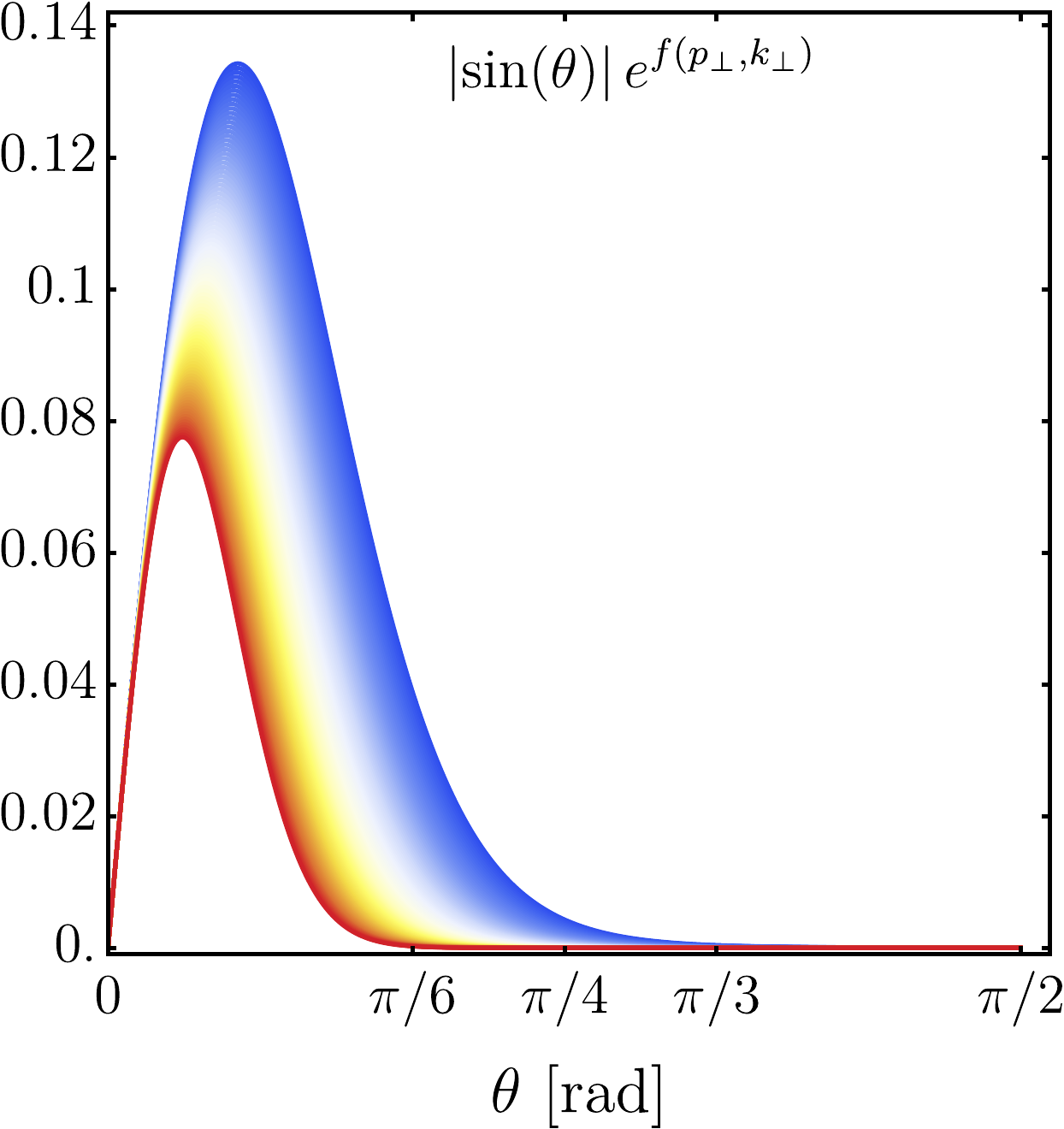}
         \caption{$\wk=1$ GeV, $\eB=5m_\pi^2$}
         \label{fig:expf_b}
     \end{subfigure}
     \hfill
     \begin{subfigure}[b]{0.31\textwidth}
         \centering
         \includegraphics[width=\textwidth]{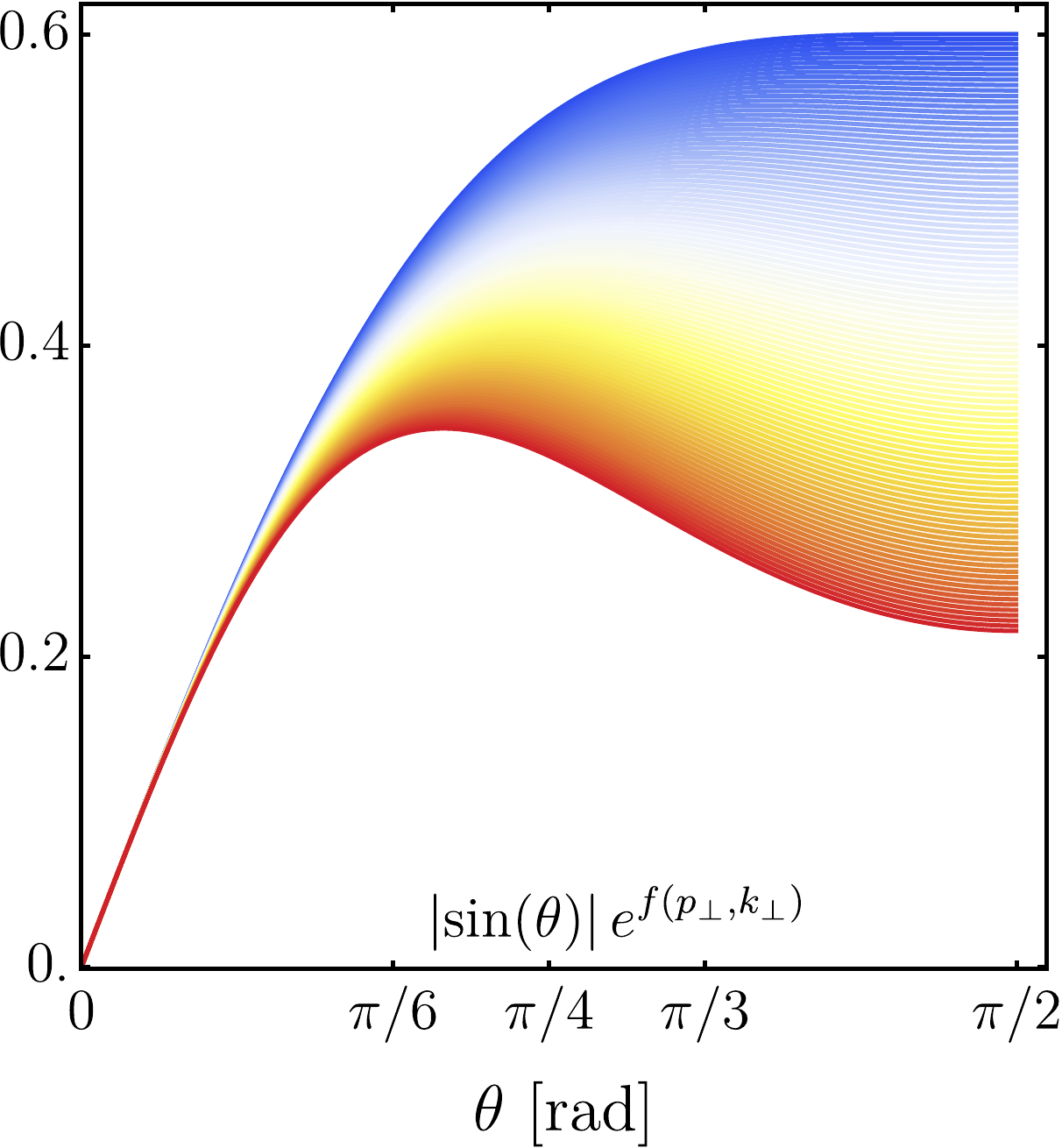}
         \caption{$\wk=0.1$ GeV, $\eB=m_\pi^2$}
         \label{fig:expf_c}
     \end{subfigure}\\
     \vspace{0.4cm}
     \includegraphics[scale=0.7]{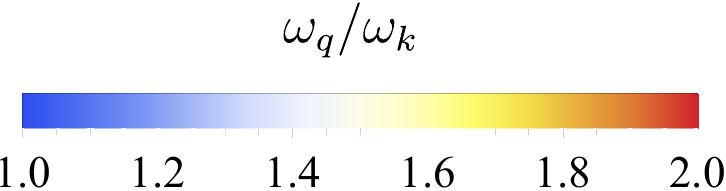}
        \caption{Angular dependence of the factor $|\qt|e^{f\left(p_{\perp}, k_{\perp}\right)}$ of Eq.~(\ref{defstildegamas}) as a function of the photon's energy $\omega_q$ and the angle with respect to the magnetic field $\theta$ for fixed values of $\wq$ and $\eB$.}
        \label{fig:expf}
\end{figure*}

To perform the integrals, we implement a Wick rotation in the $q_0$ component of the photon momentum,
namely
\begin{eqnarray}
q_0^2=\omega_q^2\to -(q_0^E)^2.
\label{wickrot0}
\end{eqnarray} 
Since the integrals to perform are written in terms of $q_\parallel^2$, we need to keep track of the consequences of this Wick rotation to then come back by means of an analytical continuation to Minkowski space.
The parallel component of the four-momentum transforms under this Wick rotation as
\begin{eqnarray}
q_\parallel^2
&\to&- \left[(q_0^E)^2 + q_3^2\right]\equiv-(q_\parallel^E)^2\nonumber\\
&=&-\omega_q^2(1+\cos^2\theta),
\label{componentsmink}
\end{eqnarray}
where $\theta$ is the angle between the photon's direction of motion and the magnetic field direction, namely the $\hat{z}$-axis. Thus, after integrating over the Feynman parameters $x$ and $y$, we can analytically continue back to Minkowski space by replacing
\begin{eqnarray}
(q_\parallel^E)^2\to \omega_q^2(1-\cos^2\theta)=\omega_q^2\sin^2\theta.
\label{backtoMink}
\end{eqnarray}

In Appendix (\ref{App2}) we calculate $\Gamma_n(\omega_q,\omega_k,q^2)$ for $n=1,2,3$ in Eq.(\ref{vertex}), by projecting Eq.(\ref{matrixelem}) onto a basis, where we have used both the transverse polarization vector and the longitudinal polarization tensor as defined in Eqs.(\ref{vperp}) and (\ref{pipar0}). The functions $\Gamma_i$ have real and imaginary parts and are defined in Eqs.(\ref{eq:Gamma1}) - (\ref{eq:Gamma3}) as
\bea
\Gamma_n&\equiv&\frac{8\pi^4 q_f g^2}{\eB}e^{f\left(p_{\perp}, k_{\perp}\right)}\left|\qt\right|\widetilde{\Gamma}_n(\omega_q,\omega_k,\theta),
\label{defstildegamas}
\eea 
where we follow through on energy conservation as $\omega_q=\omega_p+\omega_k$. Figure~\ref{fig:three graphs} shows the sum of squared amplitudes $|\widetilde{\Gamma}_n|$ as a function of the ratio of the photon energy to the quark mass for a fixed field strength and the full range of photon directions of propagation (a); again as a function of the ratio of the photon energy to the quark mass for a fixed angle of the photon propagation with respect to the magnetic field direction, for a range of field strengths (b); and as a function of the ratio of the photon energy (referred to a gluon energy) to the field strength, for a fixed value of the photon's direction of propagation, for different field intensities (c). Notice that the sum of the squared amplitudes $|\widetilde{\Gamma}_n|$ is larger for a photon propagation within the reaction plane ($\theta = \pi/2$) and that for $\theta=\pi/2$ it is also larger for smaller field intensities. On the other hand, notice that the pre-factor $|\qt|e^{f\left(p_{\perp}, k_{\perp}\right)}$ also depends on $\theta$. Figure~\ref{fig:expf} shows this dependence for different $\wk$ and $\eB$ values for a range of fixed ratios $\wq/\wk$, (a) $\wk=1$ GeV, $\eB=m_\pi^2$; (b) $\wk=1$ GeV, $\eB=5 m_\pi^2$; and (c) $\wk=0.1$ GeV, $\eB=m_\pi^2$. Notice that the pre-factor peaks at small angles for large gluon energies. In this case the squared amplitude is highly suppressed for angles close to the reaction plane. However, for small gluon energies and/or a large field strength, the pre-factor is dominated by emission angles close to the reaction plane. Since at pre-equilibrium the largest gluon abundance happens for small energies, a positive $v_2$ coefficient may be expected. Last but not least, notice that the pre-factor vanishes for $\theta=0$, which prevents photons from being emitted along the direction of the magnetic field.

\section{Summary and outlook}\label{sec4}
In this work we have studied the two-gluon one-photon vertex induced by the presence of a magnetic field. The relevant physical scenario is the pre-equilibrium stage after a heavy-ion collision where the largest field intensities are achieved. During this stage, gluons are coupled to photons by means of virtual quarks and thus the former indirectly experience the influence of the field by the interaction of this with the latter. In this sense, arguments based on the suppression of the production of electromagnetic radiation at pre-equilibrium, due to the lack of quarks ~\cite{Wang:2021oqq}, do not apply when this radiation involves processes mediated by virtual quarks. The vertex for on-shell gluons and a photon can be constructed by multiplying the longitudinal polarization tensor, that describes the polarization of two of these vector particles, times a third polarization vector which is required to have components in the transverse plane, with respect to the magnetic field, the largest energy (squared) scale. However, when the photon energy squared is allowed to be of order or larger than the magnetic field strength, this simple structure is spoiled. Nevertheless, in order to explore the bowels of this very complicated calculation, we have computed the one-loop contribution, following the strategy of Refs.~\cite{Ayala:2017vex, Ayala:2019jey}. 
This consists of explicitly computing the one-loop contribution, placing two of the loop quarks in the LLL and the other one in the 1LL. For the case when the field strength is not the largest energy scale, the computation renders structures other than the ones obtained when the field strength is the largest of the scales. Moreover, the extra structures do not respect the transversality nor the symmetry requirements for the vertex, signalling the incompleteness of the calculation. In order to avoid these shortcomings, we are currently exploring the possibility to add one-loop corrections where two and up to the three quarks in the loop occupy the 1LL. Needless to say this calculation is very challenging. However, the present result can be used to envision some of the useful properties of this vertex for calculations of photon yields and $v_2$ in a kinematical region that does not restrict the photon energy to be small compared to the field strength. For this purpose we have computed the sum of the squared amplitudes obtained when projecting the explicit computation onto the simple basis of Eq.~(\ref{vertex}). The squared amplitude shows some of the features that are to be expected, namely a dominance of photon emission along the production plane (plane transverse to the direction of the magnetic field) and a slightly larger contribution from smaller field strengths for large photon energies. Possible improvements include accounting for the contribution coming from the three loop quarks occupying the lowest and first excited Landau levels such that, still working in the large field limit, a more complete description can be achieved when the photon energy increases. This is work for the future that will be reported elsewhere.

\section*{Acknowledgments}
This work was supported in part by UNAM-DGAPA-PAPIIT Grant No. IG100322 and by Consejo Nacional de Ciencia y Tecnolog\'{\i}a Grants No. A1-S-7655 and No. A1-S-16215. J.D.C-Y. acknowledges support from Consejo Nacional de Ciencia y Tecnolog\'{\i}a Grant No. A1-S-7655 in the initial stages of this work. R. Z. acknowledges support from ANID/CONICYT FONDECYT Regular (Chile) under Grant No. 1200483. A.J.M ackowledges support from FAPESP under grant 2016/12705-7. M.E.T-Y. acknowledges support by the Simons Foundation through the Simons Foundation Emmy Noether Fellows Program at Perimeter Institute and is grateful for the hospitality of Perimeter Institute where part of this work was carried out. Research at Perimeter Institute is supported in part by the Government of Canada through the Department of Innovation, Science and Economic Development Canada and by the Province of Ontario through the Ministry of Economic Development, Job Creation and Trade.

\appendix

\begin{widetext}
\section{Calculation of the tensors $D_{i}^{\mu\nu\alpha}$}\label{App1}

The tensors $D_{i}^{\mu\nu\alpha}$ are obtained by replacing the propagators of Eqs.~(\ref{LLL}) and~(\ref{propagadores2}) into Eq.~(\ref{amplitude}), so that after integration over the coordinate space, they take the form
\bea
D_1^{\mu\nu\alpha}&=&\frac{2\ii q_f g^2}{\pi^2\eB^2}\int\frac{d^4r}{\dpi^4}\frac{d^4s}{\dpi^4}\frac{d^4t}{\dpi^4}\,\delta^{(2)}\left(\kp+\tp-\rp\right)\delta^{(2)}\left(\Sp-\pp-\rp\right)\exp\left(-\frac{r_\perp^2+s_\perp^2+t_\perp^2}{\eB}\right)\nn\\
&\times&\exp\left[\frac{2\ii}{\eB}\epsilon_{mj}(p+r-s)_m(r-t-k)_j\right]\frac{\text{Tr}\left[\gamma^1\gamma^2\gmt\slashed{r}_\parallel\gnt\slashed{t}_\parallel\gamma^\alpha\slashed{s}_\parallel\right]-2\text{Tr}\left[\gamma^1\gamma^2\gap\slashed{s}_\parallel\gamma^\mu\slashed{r}_\perp\gn\slashed{t}_\parallel\right]}{(\Sp^2-m_f^2+\ii\epsilon)(\tp^2-m_f^2+\ii\epsilon)(\rp^2-2\eB-m_f^2+\ii\epsilon)},
\label{D1ap}
\eea
\bea
D_2^{\mu\nu\alpha}&=&\frac{2\ii q_f g^2}{\pi^2\eB^2}\int\frac{d^4r}{\dpi^4}\frac{d^4s}{\dpi^4}\frac{d^4t}{\dpi^4}\,\delta^{(2)}\left(\kp+\tp-\rp\right)\delta^{(2)}\left(\Sp-\pp-\rp\right)\exp\left(-\frac{r_\perp^2+s_\perp^2+t_\perp^2}{\eB}\right)\nn\\
&\times&\exp\left[\frac{2\ii}{\eB}\epsilon_{mj}(p+r-s)_m(r-t-k)_j\right]\frac{\text{Tr}\left[\gamma^1\gamma^2\gat\slashed{s}_\parallel\gmt\slashed{r}_\parallel\gn\slashed{t}_\parallel\right]-2\text{Tr}\left[\gamma^1\gamma^2\gnp\slashed{t}_\parallel\ga\slashed{s}_\perp\gm\slashed{r}_\parallel\right]}{(\rp^2-m_f^2+\ii\epsilon)(\tp^2-m_f^2+\ii\epsilon)(\Sp^2-2\eB-m_f^2+\ii\epsilon)},
\label{D2ap}
\eea
and
\bea
D_3^{\mu\nu\alpha}&=&\frac{2\ii q_f g^2}{\pi^2\eB^2}\int\frac{d^4r}{\dpi^4}\frac{d^4s}{\dpi^4}\frac{d^4t}{\dpi^4}\,\delta^{(2)}\left(\kp+\tp-\rp\right)\delta^{(2)}\left(\Sp-\pp-\rp\right)\exp\left(-\frac{r_\perp^2+s_\perp^2+t_\perp^2}{\eB}\right)\nn\\
&\times&\exp\left[\frac{2\ii}{\eB}\epsilon_{mj}(p+r-s)_m(r-t-k)_j\right]\frac{\text{Tr}\left[\gamma^1\gamma^2\gnt\slashed{t}_\parallel\gat\slashed{s}_\parallel\gm\slashed{r}_\parallel\right]-2\text{Tr}\left[\gamma^1\gamma^2\ssl_\parallel\gmp\rs_\parallel\gn\tst\ga\right]}{(\rp^2-m_f^2+\ii\epsilon)(\Sp^2-m_f^2+\ii\epsilon)(\tp^2-2\eB-m_f^2+\ii\epsilon)}.
\label{D3ap}
\eea
\end{widetext}

The integration over the parallel momenta can be reduced to a single integral provided by the Dirac delta functions, so that
\bea
\rp&=&\tp+\kp,\nn\\
\Sp&=&\tp+\qp.
\label{ParallelMomentumConservation}
\eea

On the other hand, the integration over the perpendicular momenta can be performed by completing the square in the exponentials which leads to three Gaussian integrals over $\widetilde{r}_\perp$, $\widetilde{s}_\perp$, and $\widetilde{t}_\perp$. The latter implies the shifts over the transverse variables within the traces, given by
\begin{subequations}
\bea
t_j=\widetilde{t}_j-i\epsilon_{lj}\left(\frac{1}{2}\widetilde{r}_l+\frac{\ii}{2}\epsilon_{lm}\widetilde{r}_m-\widetilde{s}_l\right)-A_j,
\eea
\bea
s_j=\widetilde{s}_j-\frac{1}{2}\left(\ii\epsilon_{jl}\widetilde{r}_l-\widetilde{r}_j\right)-B_j,
\eea
and
\bea
r_j=\widetilde{r}_j-\frac{1}{4}C_j,
\eea
\end{subequations}
where
\begin{subequations}
\bea
A_j=i\epsilon_{lj}\left(B_l-\frac{1}{4}C_l\right),
\label{vectorA}
\eea
\bea
B_j=\frac{1}{2}\left[\frac{1}{4}C_j-p_j-\ii\epsilon_{jl}\left(k_l-\frac{1}{4}C_l\right)\right],
\label{vectorB}
\eea
and
\bea
C_j=p_j-k_j+i\epsilon_{jl}(p+k)_l.
\label{vectorC}
\eea
\end{subequations}

Note that the linear terms in the transverse tilde variables vanishes. 

For the parallel integration, we perform a Feynman parametrization over the denominators of the tensors $D_{i}^{\mu\nu\alpha}$. For example, the denominator of the tensor $D_{1}^{\mu\nu\alpha}$ can be written as:
\bea
d_1&=&\frac{1}{(\Sp^2-m_f^2+\ii\epsilon)(\tp^2-m_f^2+\ii\epsilon)(\rp^2-2\eB-m_f^2+\ii\epsilon)}\nn\\
&=&\int_0^1 dx \int_0^1dy \int_0^1dz\frac{2\delta(x+y+z-1)}{\text{den}_1^3},
\eea
where
\bea
\text{den}_1&=&x\left(s_\parallel^2-m_f^2+i\epsilon\right)+y\left(t_\parallel^2-m_f^2+\ii\epsilon\right)\nn\\
&+&z\left(r_\parallel^2-2\eB-m_f^2+\ii\epsilon\right).
\eea

Applying the momentum conservation of Eq.~(\ref{ParallelMomentumConservation}):
\bea
\text{den}_1&=&x\left[(\tp+\qp)^2-m_f^2+\ii\epsilon\right]+y\left(t_\parallel^2-m_f^2+\ii\epsilon\right)\nn\\
&+&z\left[(\tp+\kp)^2-2\eB-m_f^2+\ii\epsilon\right]\nn\\
&=&\ell_\parallel^2-\Delta_1+\ii\epsilon,
\eea
with
\bea
\lp=\tp+x\qp+z\kp\equiv\tp+\Qp,
\eea
and $\Delta_1(x,y)$ is defined in Eq.~(\ref{Delta1}). The tensor $D_{1}^{\mu\nu\alpha}$ splits into two components, correspondent to each trace, namely:
\bea
D_{1}^{\mu\nu\alpha}=D_{1\text{(a)}}^{\mu\nu\alpha}+D_{1\text{(b)}}^{\mu\nu\alpha},
\eea
so that after eliminating odd powers of $\lp$:
\begin{widetext}
\bea
&&D_{1\text{(a)}}^{\mu\nu\alpha}=\frac{4\dpi^5\ii q_f g^2}{\eB}e^{f\left(p_{\perp}, k_{\perp}\right)}\epsilon_{ij}g_\perp^{i\mu}g_\perp^{j\nu}\int_0^1 dx \int_0^{1-x}dy\int\frac{d^2\lp}{\dpi^2}\frac{1}{\left[\lp^2-\Delta_1(x,y)+\ii\epsilon\right]^3}\nn\\
&\times&\Bigg\{\left(\qp^\alpha-\Qp^\alpha-\kp^\alpha+\Qp^\alpha-\Qp^\alpha\right)\lp^2+\left(\kp-2\Qp-\qp+2\Qp+\kp+\qp-2\Qp\right)\cdot\lp\lp^\alpha\nn\\
&+&\left(\Qp^2-\kp\cdot\Qp\right)\left(\qp^\alpha-\Qp^\alpha\right)-\left(\Qp^2-\qp\cdot\Qp\right)\left(\kp^\alpha-\Qp^\alpha\right)-\left[\left(\qp-\Qp\right)\cdot\kp-\qp\cdot\Qp+\Qp^2\right]\Qp^\alpha\Bigg\},
\eea
and
\bea
&&D_{1\text{(b)}}^{\mu\nu\alpha}=-\frac{2\dpi^5\ii q_f g^2}{\eB}e^{f\left(p_{\perp}, k_{\perp}\right)}\epsilon_{ij}C^i\int_0^1 dx \int_0^{1-x}dy\int\frac{d^2\lp}{\dpi^2}\frac{1}{\left[\lp^2-\Delta_1(x,y)+\ii\epsilon\right]^3}\nn\\
&\times&\Bigg\{g_\perp^{b\nu}\left[2\lp^\mu\lp^\alpha-\Qp^\mu\left(\qp^\alpha-\Qp^\alpha\right)-\Qp^\alpha\left(\qp^\mu-\Qp^\mu\right)-\lp^2 g_\parallel^{\alpha\mu}-\left(\Qp^2-\qp\cdot\Qp\right)g_\parallel^{\alpha\mu}\right]\nn\\
&-&g_\perp^{b\mu}\left[2\lp^\nu\lp^\alpha-\Qp^\nu\left(\qp^\alpha-\Qp^\alpha\right)-\Qp^\alpha\left(\qp^\nu-\Qp^\nu\right)-\lp^2 g_\parallel^{\alpha\nu}-\left(\Qp^2-\qp\cdot\Qp\right)g_\parallel^{\alpha\nu}\right]\Bigg\}.
\eea

The integration over $\lp$ yields:
\bea
D_{1\text{(a)}}^{\mu\nu\alpha}&=&\frac{4\pi^4 q_f g^2}{\eB}e^{f\left(p_{\perp}, k_{\perp}\right)}\epsilon_{ij}g_\perp^{i\mu}g_\perp^{j\nu}\int_0^1 dx \int_0^{1-x}dy\left(\frac{1}{\Delta_1}\right)^{2}\nn\\
&\times&\Bigg\{\left(2\Qp^\alpha-\qp^\alpha\right)\Delta_1+\left(\Qp^2-\kp\cdot\Qp\right)\pp^\alpha+\left(\pp\cdot\Qp\right)\kp^\alpha-\left[\Qp^2+\left(\qp-2\Qp\right)\cdot\kp\right]\Qp^\alpha\Bigg\},
\eea
and
\bea
D_{1\text{(b)}}^{\mu\nu\alpha}&=&-\frac{8\pi^4 q_f g^2}{\eB}e^{f\left(p_{\perp}, k_{\perp}\right)}\epsilon_{ij}C^i\int_0^1 dx \int_0^{1-x}dy\left(\frac{1}{\Delta_1}\right)^{2}\nn\\
&\times&\Bigg\{g_\perp^{j\nu}\left[\left(\qp\cdot\Qp-\Qp^2\right)g_\parallel^{\alpha\mu}-\left(\Qp^\mu\left(\qp^\alpha-\Qp^\alpha\right)+\Qp^\alpha\left(\qp^\mu-\Qp^\mu\right)\right)\right]\nn\\
&-&g_\perp^{j\mu}\left[\left(\qp\cdot\Qp-\Qp^2\right)g_\parallel^{\alpha\nu}-\left(\Qp^\nu\left(\qp^\alpha-\Qp^\alpha\right)+\Qp^\alpha\left(\qp^\nu-\Qp^\nu\right)\right)\right]\Bigg\}.
\label{D1b}
\eea
\end{widetext}

Implementing that for on-shell propagation, the four-momenta are parallel, namely, Eqs.~(\ref{colinealmomenta}), the above structures can be written as
\bea
&&D_1^{\mu\nu\alpha}=\mathcal{C}_{\text{(a)}}(q)\epsilon_{ij} g_\perp^{i\mu}g_\perp^{j\nu}\qp^\alpha\nn\\
&+&\mathcal{C}_{\text{(b)}}(q)\epsilon_{ij}q^i\left[g_\perp^{j\nu}\left(g_\parallel^{\mu\alpha}+\Pi_\parallel^{\mu\alpha}\right)-g_\perp^{j\mu}\left(g_\parallel^{\nu\alpha}+\Pi_\parallel^{\nu\alpha}\right)\right],\nn\\
\eea

\section{Calculation of the vertex coefficients $\Gamma_n(\omega_q,\omega_k,q^2)$}
\label{App2}

We calculate $\Gamma_n(\omega_q,\omega_k,q^2)$ for $n=1,2,3$ in Eq.(\ref{vertex}), by projecting Eq.(\ref{matrixelem}) onto the basis
\bea
\left\{v_\perp^\mu\Pi_\parallel^{\nu\alpha},v_\perp^\nu\Pi_\parallel^{\mu\alpha}, v_\perp^\alpha\Pi_\parallel^{\mu\nu}\right\},
\label{basis}
\eea
where we have used both the transverse polarization vector and the longitudinal polarization tensor as defined in Eqs.(\ref{vperp}) and (\ref{pipar0}). For each of the three $D_i^{\mu\nu\alpha}$ structures in Eq.(\ref{matrixelem}), we get the following set of results
\bea
v_\perp^\mu \Pi_\parallel^{\nu\alpha} D_{\mu\nu\alpha}^1 &=& \frac{8\pi^4 q_f g^2}{\eB}\qp^2e^{f\left(p_{\perp}, k_{\perp}\right)}\left|\qt\right|\widetilde{C}\,\mathcal{I}_1, \nonumber \\
v_\perp^\nu \Pi_\parallel^{\mu\alpha}D_{\mu\nu\alpha}^1&=&-\frac{8\pi^4 q_f g^2}{\eB}\qp^2e^{f\left(p_{\perp}, k_{\perp}\right)}\left|\qt\right|\widetilde{C}\,\mathcal{I}_1, \nonumber \\
v_\perp^\alpha \Pi_\parallel^{\mu\nu}D_{\mu\nu\alpha}^1&=&0,
\label{be1}
\eea
\bea
v_\perp^\mu \Pi_\parallel^{\nu\alpha} D_{\mu\nu\alpha}^2&=&-\frac{8\pi^4 q_f g^2}{\eB}\qp^2e^{f\left(p_{\perp}, k_{\perp}\right)}\left|\qt\right|\widetilde{B}\,\mathcal{I}_2, \nonumber \\
v_\perp^\nu \Pi_\parallel^{\mu\alpha}D_{\mu\nu\alpha}^2&=&0, \nonumber \\
v_\perp^\alpha \Pi_\parallel^{\mu\nu}D_{\mu\nu\alpha}^2&=&\frac{8\pi^4 q_f g^2}{\eB}\qp^2e^{f\left(p_{\perp}, k_{\perp}\right)}\left|\qt\right|\widetilde{B}\,\mathcal{I}_2,
\label{be2}
\eea
\bea
v_\perp^\mu \Pi_\parallel^{\nu\alpha} D_{\mu\nu\alpha}^3&=&0, \nonumber \\
v_\perp^\nu \Pi_\parallel^{\mu\alpha}D_{\mu\nu\alpha}^3&=&\frac{8\pi^4 q_f g^2}{\eB}\qp^2e^{f\left(p_{\perp}, k_{\perp}\right)}\left|\qt\right|\widetilde{A}_2\,\mathcal{I}_3, \nonumber \\
v_\perp^\alpha \Pi_\parallel^{\mu\nu}D_{\mu\nu\alpha}^3 &=& -\frac{8\pi^4 q_f g^2}{\eB}\qp^2e^{f\left(p_{\perp}, k_{\perp}\right)}\left|\qt\right|\widetilde{A}_2\,\mathcal{I}_3.
\label{be3}
\eea

Using these results, we can go back to Eq.~(\ref{vertex}) and collect the contributions that correspond to each coefficient $\Gamma_n(\omega_q,\omega_k,q^2)$. The projection with $v_\perp^\mu \Pi_\parallel^{\nu\alpha}$ from the three sets in (\ref{be1})-(\ref{be3}), contributes to 
\bea
\Gamma_1&=&\frac{8\pi^4 q_f g^2}{\eB}e^{f\left(p_{\perp}, k_{\perp}\right)}\left|\qt\right|\qp^2\left(\widetilde{C}\mathcal{I}_1-\widetilde{B}\mathcal{I}_2\right)\nn\\
&\equiv&\frac{8\pi^4 q_f g^2}{\eB}e^{f\left(p_{\perp}, k_{\perp}\right)}\left|\qt\right|\widetilde{\Gamma}_1(\omega_p,\omega_q,\theta).
\label{eq:Gamma1}
\eea
The projection with $v_\perp^\nu \Pi_\parallel^{\mu\alpha}$ from the  three sets in Eqs.~(\ref{be1})--(\ref{be3}), contributes to 
\bea
\Gamma_2&=&\frac{8\pi^4 q_f g^2}{\eB}e^{f\left(p_{\perp}, k_{\perp}\right)}\left|\qt\right|\qp^2\left(\widetilde{A}_2\mathcal{I}_3-\widetilde{C}\mathcal{I}_1\right)\nn\\
&\equiv&\frac{8\pi^4 q_f g^2}{\eB}e^{f\left(p_{\perp}, k_{\perp}\right)}\left|\qt\right|\widetilde{\Gamma}_2(\omega_p,\omega_q,\theta),
\label{eq:Gamma2}
\eea
and simmilarly for the projection with $v_\perp^\alpha \Pi_\parallel^{\mu\nu}$, that contributes to 
\bea
\Gamma_3&=&\frac{8\pi^4 q_f g^2}{\eB}e^{f\left(p_{\perp}, k_{\perp}\right)}\left|\qt\right|\qp^2\left(\widetilde{B}\mathcal{I}_2-\widetilde{A}_2\mathcal{I}_3\right)\nn\\
&\equiv&\frac{8\pi^4 q_f g^2}{\eB}e^{f\left(p_{\perp}, k_{\perp}\right)}\left|\qt\right|\widetilde{\Gamma}_3(\omega_p,\omega_q,\theta).
\label{eq:Gamma3}
\eea

\bibliography{bibliography}

\end{document}